\documentclass[letterpaper]{article} % DO NOT CHANGE THIS
\usepackage{aaai2026}  % DO NOT CHANGE THIS
\nocopyright
\usepackage{times}  % DO NOT CHANGE THIS
\usepackage{helvet}  % DO NOT CHANGE THIS
\usepackage{courier}  % DO NOT CHANGE THIS
\usepackage[hyphens]{url}  % DO NOT CHANGE THIS
\usepackage{graphicx} % DO NOT CHANGE THIS
\usepackage{natbib}  % DO NOT CHANGE THIS AND DO NOT ADD ANY OPTIONS TO IT
\usepackage{caption} % DO NOT CHANGE THIS AND DO NOT ADD ANY OPTIONS TO IT
\usepackage{algorithm}

\usepackage{amsmath} 

\usepackage{booktabs}
\usepackage{xspace}

\newcommand{\framework}{\textit{ALA}\xspace}
\usepackage{newfloat}
\usepackage{listings}
\usepackage{multirow}

\usepackage{algpseudocode}

\DeclareCaptionStyle{ruled}{labelfont=normalfont,labelsep=colon,strut=off} % DO NOT CHANGE THIS
\floatstyle{ruled}
\newfloat{listing}{tb}{lst}{}
\floatname{listing}{Listing}
\title{Selection-Based Vulnerabilities: Clean-Label Backdoor Attacks in Active Learning}
\author {
    Yuhan Zhi\textsuperscript{\rm 1},
    Longtian Wang\textsuperscript{\rm 1},
    Xiaofei Xie\textsuperscript{\rm 2},
    Chao Shen\textsuperscript{\rm 1},
    Qiang Hu\textsuperscript{\rm 3},
    Xiaohong Guan\textsuperscript{\rm 1}
}
\affiliations {
    \textsuperscript{\rm 1}Xi'an Jiaotong University\\
    \textsuperscript{\rm 2}Singapore Management University\\
    \textsuperscript{\rm 3}Tianjin University\\
    zyh1123@stu.xjtu.edu.cn
}
\usepackage{bibentry}
\begin{document}

\maketitle

\begin{abstract}

Active learning~(AL), which serves as the representative label-efficient learning paradigm, has been widely applied in resource-constrained scenarios. The achievement of AL is attributed to acquisition functions, which are designed for identifying the most important data to label. Despite this success, one question remains unanswered: \textit{is AL safe?} In this work, we introduce \framework, a practical and the first framework to utilize the acquisition function as the poisoning attack surface to reveal the weakness of active learning. Specifically, \framework optimizes imperceptibly poisoned inputs to exhibit high uncertainty scores, increasing their probability of being selected by acquisition functions. To evaluate \framework, we conduct extensive experiments across three datasets, three acquisition functions, and two types of clean-label backdoor triggers. Results show that our attack can achieve high success rates (up to 94\%) even under low poisoning budgets (0.5\%–1.0\%) while preserving model utility and remaining undetectable to human annotators. Our findings remind active learning users: acquisition functions can be easily exploited, and active learning should be deployed with caution in trusted data scenarios.

% Active learning (AL) has become a powerful paradigm for reducing labeling costs by selectively querying the most informative samples from an unlabeled data pool. While the efficiency of AL largely relies on its data selection strategy, the security implications of such selection mechanisms remain underexplored.
% In this paper, we present the first study to demonstrate that the data selection process itself can serve as an attack surface for backdoor injection in active learning.
% We show that existing clean-label backdoor attacks—originally developed for supervised learning—can be effectively adapted to AL settings by leveraging the model's selection mechanism. We optimize imperceptibly poisoned inputs to exhibit high uncertainty scores, increasing their likelihood of being selected for labeling and subsequently injected into the training set. Our method requires no label manipulation and remains stealthy during annotation, yet successfully implants backdoors via standard AL pipelines.
% We conduct extensive experiments across four datasets, multiple selection methods, and three types of clean-label backdoor triggers. Results show that our attack achieves high success rates even under low poisoning budgets (0.1\%–0.5\%) while preserving model utility and remaining undetectable to human annotators. Our findings highlight a critical paradox: the selection strategies that drive AL's efficiency can also expose it to backdoor vulnerabilities, especially in real-world systems that learn from evolving data distributions.

\end{abstract}

\section{Introduction}
%AL popular usage and introduction
%selection method 本身可能是安全漏洞
%motivation setting
%brief introduction of  method

Active learning (AL) is a learning paradigm designed to reduce annotation costs by iteratively selecting and labeling the most informative samples from an unlabeled data pool as training data~\cite{settles2009active, munjal2022towards, ren2021survey}. Acquisition function, which identifies the data to be labeled, is the core of AL. Multiple acquisition functions have been proposed~\cite{tong2001support, joshi2009multi, li2013adaptive,beluch2018power}, where uncertainty based ones are the most representative, such as prediction confidence-based function and entropy score-based function~\cite{wang2014new}
. With the increasingly proposed effective acquisition functions, AL has been more and more useful, especially when labeling is expensive, enabling the model to achieve comparable accuracy with fewer labeled samples.

% Rather than labeling the entire dataset, AL relies on a selection strategy to identify samples that are expected to provide the greatest benefit when added to the training set.  
% We intentionally use the term \textit{selection strategy}—rather than the more common \textit{query strategy}—to emphasize that our focus is on the sample selection mechanism itself. 
% This approach is especially useful when labeling is expensive, enabling the model to achieve comparable accuracy with fewer labeled examples.

In real-world applications, AL is frequently deployed in two scenarios: 1) training deep learning (DL) models from scratch with limited data labeling budgets~\cite{DBLP:conf/iclr/SenerS18,wang2014new}, and 2) continual learning for quickly new data distribution adaptation~\cite{DBLP:conf/icse/KimFY19,hu2024active}. For the latter one, when a model has been deployed in the wild, it is common that the new unseen data samples follow different data distributions, i.e., out-of-distribution (OOD) data, from the original training data. In this situation, AL is used to select fine-tuning data for model enhancement. Such settings arise in medical diagnosis, autonomous systems, and data-centric AI applications~\cite{DBLP:conf/explimed/SantosC24, DBLP:journals/corr/abs-2403-02877}, where models must incrementally learn new concepts or domains from previously unseen data.

% In real-world applications, AL is frequently deployed not only for initial model training but also for continual adaptation, where a model trained on in-distribution (ID) data must extend its performance to a stream of previously unseen, out-of-distribution (OOD) data.  
% Such settings arise in medical diagnosis, autonomous systems, and data-centric AI applications, where models must incrementally learn new concepts or domains from previously unseen data.

Even though AL has achieved great success in multiple domains and scenarios, the safety of AL has rarely been discussed. As the heart of AL, the acquisition function determines which unlabeled samples should be annotated. In this manner, one assumption is -- the potential selection bias in acquisition function has the potential to be exploited to inject poisoned samples into the training set, thus, serves as the attack surface.

% At the heart of AL lies the selection strategy, which determines which unlabeled samples should be annotated at each iteration.  
% This process directly shapes the model's learning trajectory and efficiency.  
% Despite its central role, prior work has rarely considered that the selection strategy itself—while designed to improve learning—may introduce new security vulnerabilities.  

To this end, in this paper, we introduce \framework, the first framework to utilize acquisition functions to reveal the weakness of active learning from the poisoning attack perspective. We consider a white-box threat model where attackers can access the model parameters. \framework strategically injects clean-label poisoned samples and exploits the construction design of acquisition functions~(selecting data samples with high uncertainty) to increase their probability of being included in the training set. Specifically, \framework first selects samples that are near the decision boundaries~(quantified by the high uncertainty scores) as seed data. Then, \framework applies poisoning attack methods with a selection-aware optimization algorithm to generate poisoned samples while further increase the uncertainty scores of these generated samples. In this way, the poisoned samples are more likely to be selected by acquisition functions and therefore, injected into the training set. 

To evaluate \framework, we conduct comprehensive experiments on three datasets (Fashion-MNIST, CIFAR-10, and SVHN) with two clean-label backdoor triggers (CL~\cite{turner2019label} and SIG~\cite{barni2019new}) and three acquisition functions (Entropy, Margin, and Least Confidence~\cite{wang2014new}). %Experimental results demonstrate that \framework injects 100\% poisoned samples into the training set when the poison ratio is 0.5\%, which is 88\% higher than the Random selection baseline. 
Our results show that poisoned samples injected by \framework can lead to attack success rates of up to 94\% on the AL-trained model under uncertainty-based acquisition functions, even with a low poisoning budget, achieving a 43\% improvement over the Random selection baseline, and highlighting the potential vulnerability of AL in practical scenarios.
Moreover, we find that in AL scenarios, the SIG trigger is substantially more effective and robust than CL for clean-label backdoor injection, consistently achieving higher ASRs across datasets and acquisition functions.

\paragraph{Our Contributions.}  
\begin{itemize}
  \item We identify the acquisition function in AL as a new and realistic attack surface for clean-label backdoor injection.
  \item We propose \framework, the first framework to attack active learning with a selection-aware optimization strategy that aligns poisoned inputs with the model’s selection preferences to enhance attack success.
  \item We conduct extensive experiments across three datasets, three acquisition functions, and two clean-label attack types, demonstrating strong attack effectiveness under multiple common AL configurations, revealing a potential real-world vulnerability of active learning.
\end{itemize}

Our results raise important concerns about the security of AL pipelines, and highlight the need to incorporate selection-aware threat models into the design of robust active learning systems.

\section{Background}
%AL framework (data pool, selection strategy, iteration)
%selection method (especially uncertainty-based)
%backdoor attack (especially clean label)

\subsection{Active Learning}
Active learning is a learning paradigm that aims to reduce annotation costs by selectively querying labels for only the most informative samples~\cite{settles2009active, munjal2022towards}. Rather than training on a large fully labeled dataset, an AL algorithm maintains a small labeled set $\mathcal{L}$ and a large unlabeled pool $\mathcal{U}$, and iteratively selects samples from $\mathcal{U}$ to be labeled by an oracle (e.g., a human annotator)~\cite{ren2021survey}. The newly labeled samples are added to $\mathcal{L}$, and the model is retrained. This loop continues until a predefined labeling budget is exhausted or model performance converges.

The AL framework typically includes three components: (1) a model or learner, (2) an acquisition function for determining which samples to label next, and (3) a labeling oracle. This iterative loop has been widely applied in scenarios such as medical imaging, autonomous driving, and real-world systems where labeling is expensive and continuous adaptation is necessary.

\subsection{Acquisition Functions in Active Learning}
The effectiveness of AL heavily depends on the design of its acquisition function, which determines which unlabeled samples are most beneficial to label at each epoch. Among various approaches, uncertainty-based acquisition functions~\cite{tong2001support, joshi2009multi, li2013adaptive,beluch2018power} are the most widely adopted due to their simplicity and effectiveness. These methods prioritize samples on which the current model is most uncertain.

Common uncertainty-based acquisition functions include:
\begin{itemize}
  \item \textbf{Entropy}: Selects samples with the highest entropy in the predicted class distribution. For a sample $x$, let $p(y \mid x)$ be the model’s predicted probability distribution over $K$ classes. The entropy~\cite{shannon1948mathematical} is computed as:
\[
\mathcal{H}(x) = - \sum_{k=1}^{K} p(y_k \mid x) \log p(y_k \mid x).
\]
Samples with higher entropy indicate greater prediction uncertainty.

\item \textbf{Margin}: Selects samples with the smallest margin between the top two class probabilities~\cite{roy2001toward}.
  \item \textbf{Least Confidence}: Selects samples with the lowest maximum predicted probability~\cite{wang2014new}.
  
\end{itemize}

Other acquisition functions target alternative objectives beyond uncertainty.
Diversity-based methods aim to select samples that are both representative of the input distribution and mutually diverse, in order to cover different regions of the feature space~\cite{sener2017active, bilgic2009link, gal2017deep}.
Hybrid methods combine multiple selection signals—typically blending uncertainty with diversity—to improve robustness and sample efficiency~\cite{ash2019deep, shui2020deep, yin2017deep}.
% Other selection strategies focus on different objectives:
% Diversity-based: Select samples that are representative and diverse~\cite{sener2017active, bilgic2009link, gal2017deep}. 
%   %(e.g., core-set).
% Hybrid: Combine multiple signals such as uncertainty and diversity~\cite{ash2019deep, shui2020deep, yin2017deep}. 
  %For example, BADGE~\cite{ash2019deep} selects uncertain samples that also induce diverse gradient embeddings.

In this work, we focus on \textbf{uncertainty-based acquisition functions}, as they are generally effective in practice and are commonly used in AL~\cite{li2022empirical, li2024survey}.

\subsection{Backdoor Attacks and Clean-Label Threats}
Backdoor attacks~\cite{gu2017badnets} are a class of data poisoning attacks that aim to implant hidden malicious behavior into machine learning models. Typically, the attacker injects a small number of specially crafted training samples—known as poisoned samples—that contain a specific trigger pattern. These samples are labeled as the attacker's desired target class. At inference time, any input containing the same trigger will be misclassified into the target class, while the model maintains high accuracy on clean inputs.

Traditional backdoor attacks often assume that the attacker can manipulate both the training data and the associated labels~\cite{liu2018trojaning, zhong2020backdoor, ji2017backdoor}. However, this assumption may not hold in many real-world scenarios, especially when labels are provided by human annotators or trusted pipelines~\cite{turner2019label}. To address this limitation, clean-label backdoor attacks have emerged as a more realistic and stealthy threat model by preserving the ground-truth labels of poisoned samples~\cite{saha2020hidden}.

Clean-label backdoor attacks exploit the mismatch between human and model perception by embedding imperceptible triggers into inputs without altering their ground-truth labels. While human annotators assign correct labels based on visual semantics, the model may learn to associate subtle patterns—such as high-frequency or localized signals—with a specific target class. As a result, inputs containing the same trigger at test time can be misclassified into the target class, even if they are semantically unrelated.
Representative techniques include:

\begin{itemize}
  \item \textbf{CL}: Adds localized, low-visibility triggers (e.g., small patches) without altering the sample's semantics~\cite{turner2019label}.
  \item \textbf{SIG}: Applies imperceptible, frequency-domain perturbations as triggers~\cite{barni2019new}.
  %\item \textbf{F-Trojan}: Embeds triggers into intermediate feature representations while maintaining clean labels~\cite{saha2020hidden}.
\end{itemize}

These techniques are particularly effective in settings where the labeling process is external and cannot be manipulated. In our work, we adopt and adapt these clean-label backdoor methods to the context of active learning, where the attacker cannot flip labels and must remain undetected by human annotators.

\section{Methodology}
\subsection{Problem setting}

We consider a realistic and common deployment scenario: a deployed model has already been well trained on an initial labeled dataset and achieves strong performance. Given newly collected unlabeled data that follow a different distribution~(OOD data) from the original training data, we employ AL to conduct distribution adaptation for the pre-trained model under a specific labeling budget.

% Fighting
% We consider a realistic and increasingly common deployment scenario: an active learning model that has already been well trained on an initial labeled dataset and achieves strong performance on its original ID data. The goal is to extend this model to a stream of newly collected OOD data via AL-based selective labeling.

We focus on this setting for both technical and practical reasons. From a technical standpoint, poisoning the initial training set is often infeasible, as it typically requires insider access to the original data pipeline. In contrast, newly arriving OOD data is more accessible and provides a natural point of interaction for external attackers. In practice, attacking a poorly trained model is of limited value, as its predictions are already unreliable. In contrast, attacking a deployed, well-performed model makes real-world impacts and therefore should be more carefully considered.

% Targeting a well-performing model during its continual adaptation phase ensures that any injected backdoor will have measurable and impactful consequences.

In our setting, the AL model follows a standard loop: at each epoch, the model selects a batch of unlabeled OOD samples using a predefined acquisition function~(e.g., Entropy, Margin, Least Confidence), obtains their labels through human annotation (i.e., oracle in our experiments), and retrains on the updated labeled set. This process repeats until the labeling budget is exhausted or target performance is reached.

Our attack aims to inject a clean-label backdoor during this OOD adaptation process. Specifically, we poison a small number of OOD samples by embedding an imperceptible trigger, such that the final model misclassifies any input containing the trigger into a specific target class, while maintaining high accuracy on clean inputs from both ID and OOD distributions.

To make the threat model explicit, we assume the following capabilities and constraints for the attacker:

\begin{itemize}
    \item \textbf{Model access}: The attacker has access to the current AL model before the selection step on incoming OOD data. This includes the architecture and parameters of the model.
    
    \item \textbf{Data injection}: The attacker can access and modify a subset of the newly arriving OOD samples \textit{before they are injected into the unlabeled pool}. However, the attacker has \textbf{no access} to the original ID training data.
    
    \item \textbf{Label integrity}: The attacker cannot alter the ground-truth labels, nor interfere with the annotation process, which is assumed to be performed by human annotators.
    
    % \item \textbf{Stealth constraints}: To avoid detection and ensure correct labeling, the attacker cannot apply noticeable or semantically disruptive modifications. %For example, trimming a digit `7' to resemble a `1' in MNIST would violate the clean-label assumption.
\end{itemize}

These assumptions reflect realistic attack surfaces in deployed AL systems. Meanwhile, this setting introduces unique challenges: the attacker must craft poisoned samples that satisfy the clean-label constraint, and yet ensure that they are selected by the AL agent despite being rare in a large, unfamiliar OOD pool.

\subsection{Overview of \framework}

\begin{figure}[t]
    \centering
    \includegraphics[width=0.9\linewidth]{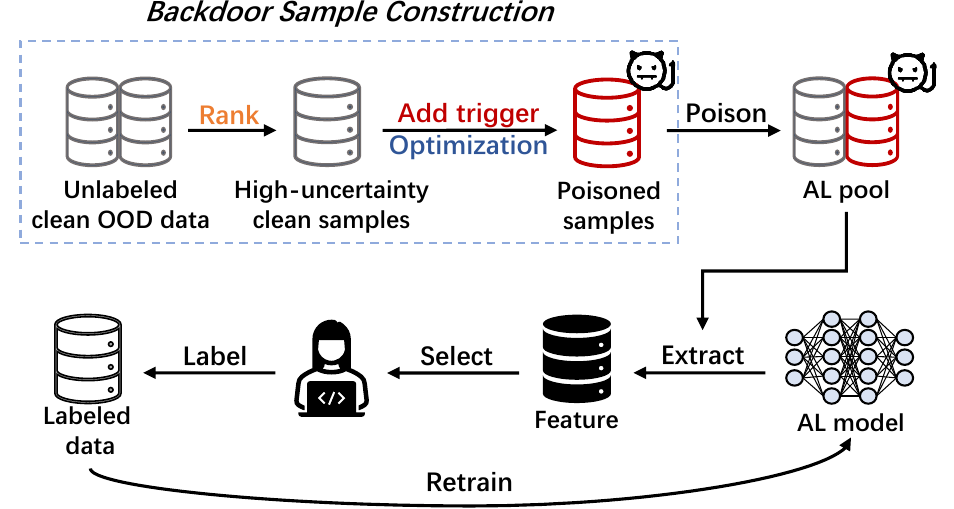}
    \caption{Overview of \framework.}
    \label{fig:overview}
\end{figure}

Figure~\ref{fig:overview} illustrates the overall workflow of \framework. It first uses the current-epoch AL model~(or pre-trained model) to rank the incoming OOD data based on their uncertainty scores. We then identify high-uncertainty samples from the attack target class. After that, according to a predefined poisoning ratio, \framework embeds an invisible trigger to the top uncertain samples. Since adding the trigger may change the prediction confidence of the model on the data, in turn, affects the uncertainty scores, we design a selection-aware optimization algorithm to increase the attack success rate and the uncertainty of the poisoned data.

% genetic algorithm to optimize each sample to increase its uncertainty.

% overall process of our attack and its position within the AL loop. We consider an AL model that has already achieved good performance on its current ID dataset. The attack is launched before a batch of newly collected unlabeled OOD data is injected into the AL pool.

% We first use the current-round AL model to rank the incoming OOD data. Among these, we identify high-uncertainty samples from the attack target class. According to a predefined poisoning ratio, we select the required number of such samples and embed an invisible trigger into them. Since adding the trigger may change the sample’s uncertainty score, we apply a genetic algorithm to individually optimize each sample to increase its uncertainty.

The optimized poisoned samples are then injected into the candidate pool under acquisition. The AL agent selects samples based on their extracted features, obtains labels from the oracle, and retrains the AL model. This process is repeated throughout the AL loop, allowing poisoned samples to be gradually incorporated and enabling the model to learn the backdoor behavior over time.

\begin{algorithm}[t]
\caption{Backdoor Sample Construction}
\label{alg:backdoor}
\begin{algorithmic}[1]
\Require Unlabeled pool $\mathcal{U}$, target class $y_t$, poisoning ratio $\rho$, AL model $M$, trigger $T$, GA iterations $G$, population size $P$, Entropy calculation $H$
\Ensure Poisoned sample set $\mathcal{P}$

\State $\mathcal{P} \gets \emptyset$, $\mathcal{C} \gets \emptyset$ \Comment{Initialize}
\State Rank $\mathcal{U}$ by entropy $H(M(x))$ in descending order
\For{top-ranked $x \in \mathcal{U}$}
    \State $y \gets \text{QueryLabel}(x)$  \Comment{Human annotation}
    \If{$y = y_t$}
        \State $\mathcal{C} \gets \mathcal{C} \cup \{x\}$
        \If{$|\mathcal{C}| \geq \rho \cdot |\mathcal{U}|$} \textbf{break} \EndIf
    \EndIf
\EndFor

\For{$x \in \mathcal{C}$}
    \State $x_{\text{best}} \gets x$, $H_{\text{max}} \gets -\infty$
    \For{$g = 1$ to $G$}
        \State Generate mutants $\{m_1, \dots, m_P\}$ from $x_{\text{best}}$ via augmentation
        \State $\mathcal{P}_{\text{cand}} \gets \{\text{ApplyTrigger}(m_i, T)\}_{i=1}^P$
        \State $index \gets \arg\max_{p \in \mathcal{P}_{\text{cand}}} H(M(p))$
        \State $p^* \gets m_{index}$
        \If{$H(M(p^*)) > H_{\text{max}}$}
            \State $x_{\text{best}} \gets p^*$, $H_{\text{max}} \gets H(M(p^*))$
        \EndIf
        \State $x \gets \text{RemoveTrigger}(p^*, T)$  \Comment{Clean seed for next iter}
    \EndFor
    \State $\mathcal{P} \gets \mathcal{P} \cup \{x_{\text{best}}\}$
\EndFor
\State \Return $\mathcal{P}$
\end{algorithmic}
\end{algorithm}

\subsection{Backdoor Sample Construction}

The backdoor sample construction process is the core of \framework, which contains two key steps: target-class candidate selection and selection-aware optimization. Algorithm~\ref{alg:backdoor} shows the detailed process.

% This section describes how poisoned samples are constructed such that they both satisfy the clean-label constraint and have a high probability of being selected by the AL agent.

\paragraph{Target-class candidate selection.}
Clean-label backdoor attacks target a specific class $y_t$, and poisoning is only applied to samples belonging to this class. Since the incoming OOD samples are unlabeled, the attacker must manually identify which samples belong to $y_t$. Given this setting, we aim to minimize labeling cost by first ranking the unlabeled OOD pool $\mathcal{U}$ using the current AL model’s uncertainty score, in this work we use entropy, though any uncertainty-based metric could be used (Line 2), and then sequentially querying the ground-truth labels of the top-ranked samples, which in practice would require human annotation (Line 3-6). The process stops once $\rho \cdot |\mathcal{U}|$ samples from class $y_t$ are found (Line 7), where $\rho$ is the predefined poisoning ratio. This strategy ensures that the selected samples are both label-efficient and more likely to be chosen by the AL agent.

\paragraph{Selection-aware optimization.}

To perform the poisoning attack, a trigger needs to be embedded in the data. However, adding a trigger can affect the uncertainty of a sample, potentially reducing the chance that it will be queried by the uncertainty-based acquisition function. To mitigate this, we design a selection-aware optimization procedure based on a genetic algorithm (GA) to increase each poisoned sample’s uncertainty score~(i.e., entropy). Specifically, given a clean sample as a seed, \framework first mutates it and generates multiple mutants (Line 14). Then, the poisoning attack is used to inject triggers into these mutants (Line 15). After that, we calculate the uncertainty score of each poisoned sample and keep the one with the highest uncertainty (Line 16-19). After removing its trigger, this sample will be used as the seed in the next iteration (Line 20). Finally, the poisoned sample with the highest entropy in the final population is selected as the optimized poisoned sample (Line 22).

% Adding a trigger can affect the uncertainty of a sample, potentially reducing the chance that it will be queried by the acquisition function. To mitigate this, we design a selection-aware optimization procedure to increase each poisoned sample’s uncertainty score~(i.e., entropy). Specifically, we use a genetic algorithm (GA) to independently evolve each poisoned sample over multiple generations. The GA applies mutation and crossover operations to generate candidate variants and selects the ones with the highest entropy for the next generation. This process continues until a maximum number of generations is reached. Finally, the sample with the highest entropy in the final population is selected as the optimized poisoned sample.

% \paragraph{Trigger injection.}
% An imperceptible trigger $\delta$ is embedded into each selected sample.

We adopt existing clean-label attack methods—CL and SIG—for trigger construction. To remain stealthy, the trigger is designed to be visually imperceptible and does not alter the semantic content of the image. The ground-truth label remains unchanged, which satisfies the clean-label constraint and avoids suspicion during the annotation process.

\section{Experiments}
\subsection{Datasets and Models}
We conduct experiments on three widely used benchmark datasets: Fashion-MNIST~\cite{xiao2017fashion}, a grayscale image dataset of clothing items with 10 classes; CIFAR-10~\cite{krizhevsky2009learning}, a 10-class natural image dataset with low-resolution color images; and SVHN~\cite{netzer2011reading}, a real-world digit classification dataset derived from street view house numbers. These datasets cover a range of visual domains and classification difficulties.

For each dataset, we use a commonly adopted architecture with sufficient capacity for the corresponding task. Specifically, we use LeNet-5~\cite{lecun1998gradient} for Fashion-MNIST, ResNet-18~\cite{he2016deep} for CIFAR-10, and MobileNetV2~\cite{sandler2018mobilenetv2} for SVHN.

\subsection{Setup}
%AL setting, backdoor setting, optimization setting, repetition, data distribution (ID ood)
\paragraph{AL Setting.}
Following previous work~\cite{hu2024active}, we simulate an iterative AL loop with a total labeling budget of 10\% of the unlabeled data. At each epoch, the model selects 1\% of samples from the unlabeled pool using a predefined acquisition function. We evaluate three widely used uncertainty-based acquisition functions: Entropy, Margin, and Least Confidence. Following standard AL practice, the model is incrementally trained at each epoch using the updated labeled set, continuing from the model parameters obtained in the previous epoch.

\paragraph{Data Distribution.}
To simulate continuous learning in deployment time on unfamiliar data, we construct an unlabeled OOD pool by applying a randomly sampled corruption from CIFAR-10-C~\cite{hendrycks2019benchmarking} to each test image, with a randomly chosen severity level. CIFAR-10-C provides 19 types of common visual corruptions that can significantly degrade model performance, each with five severity levels.
A corrupted sample is added to the OOD pool if it is misclassified by the pretrained AL model. Otherwise, the corruption process is repeated until a misclassified variant is found. This procedure ensures that the OOD pool consists of visually plausible samples that the model fails to recognize, reflecting realistic data drift during post-deployment learning. The ID data used for pretraining the AL model corresponds to the original training split of each dataset.

\paragraph{Backdoor Attack Configuration.}
The current version of \framework integrates two representative clean-label backdoor attack methods. \framework is extensible, and any new attack methods can be easily employed in it.

\begin{itemize}
    \item \textbf{Clean-Label (CL)}~\cite{turner2019label}: Adversarial samples are generated via projected gradient descent (PGD) attack. The perturbation is constrained under an $\ell_\infty$ bound of $\epsilon = 32$.
    \item \textbf{SIG}~\cite{barni2019new}: An imperceptible sinusoidal signal trigger is overlaid onto each poisoned sample. The signal is defined by a spatial frequency $f = 6$ and amplitude $\delta = 50$, resulting in a smooth, high-frequency pattern that is visually indistinguishable to humans but learnable by the model.
\end{itemize}
All poisoned samples are assigned their correct ground-truth labels and visually resemble benign data. We evaluate two poisoning ratios: 0.5\% and 1.0\% of the OOD pool. Poisoning is applied class-wise for each of the 10 classes, and results are averaged over classes.

\paragraph{Selection-Aware Optimization.}
To increase the likelihood that poisoned samples are selected by the AL agent, we apply a GA to optimize their uncertainty (measured via entropy). For each poisoned sample, the GA is run independently with a population size of 100, tournament size of 5, and mutation rate of 0.5. Crossover is performed by randomly inheriting from one parent. The mutation operations include: (1) pixel-level noise (Gaussian, salt-and-pepper, multiplicative), (2) blurring (Gaussian, uniform, median, bilateral), and (3) global transformations (brightness and contrast adjustments). We evaluate the effect of optimization at multiple checkpoints: 0, 5, 10, and 15 iterations.

%\todo{repetition, cpu gpu}
\paragraph{Experiment Configurations.}
Each active learning process is repeated three times with different random seeds, and we report the average value of all evaluation metrics.  
All experiments are conducted on a server equipped with an Intel(R) Xeon(R) Gold 6226R CPU (64 cores, 2.90GHz) and eight NVIDIA RTX 3090 GPUs.

\subsection{Evaluation Metrics}
%ASR ID-ACC OOD-ACC select_num
We evaluate our attack effectiveness and its impact on model performance using the following metrics:

\begin{itemize}
% \item \textbf{Poisoned Selection Count ($N_{select}$).}
% We report the cumulative number of poisoned samples selected by the AL agent throughout the entire active learning process.  
% Since our attack is designed to make poisoned samples more attractive to selection strategies (e.g., by increasing uncertainty), we expect $N_{select}$ to be higher when using a selection strategy compared to random sampling.  
% This metric directly evaluates whether poisoned samples are consistently prioritized by the selection mechanism, thereby quantifying the effectiveness of targeting the selection strategy itself.
\item \textbf{Poisoned Selection Rate ($R_{select}$).}
We report the percentage of poisoned samples that are selected by the AL agent during the entire active learning process.  
%Since our attack aims to make poisoned samples more attractive to acquisition functions, a higher $R_{select}$ under selection-based acquisition functions---compared to Random sampling---indicates a more effective attack.  
A higher $R_{select}$ under uncertainty-based acquisition functions—compared to Random—indicates a more effective attack
This metric directly quantifies how well the attack succeeds in manipulating the acquisition function to prioritize poisoned inputs.

\item \textbf{Attack Success Rate (ASR).}
ASR measures the fraction of test-time inputs embedded with the backdoor trigger that are misclassified into the attacker-specified target class. A higher ASR indicates a more effective backdoor attack.

\item \textbf{ID Accuracy ($Acc_{ID}$).}
To assess whether the backdoor affects the model’s performance on previously learned data, we report the classification accuracy on clean, ID test data. High $Acc_{ID}$ indicates that the attack preserves performance on the original distribution.

\item \textbf{OOD Accuracy ($Acc_{OOD}$).}
We also measure the model’s accuracy on clean (i.e., non-poisoned) OOD test data to assess whether the attack compromises generalization to newly learned classes. Maintaining high $Acc_{OOD}$ is important for stealthiness, as a noticeable drop may expose the presence of poisoned data.

\end{itemize}

\subsection{Results}
%trigger 可视化（展示其不可见性）
%ablation study：不使用优化时的对比，优化轮次不同时的对比
%不同后门攻击方法的对比
%不同攻击类别的对比
%投毒率的对比 0.5%，1%，2%
\paragraph{Effectiveness of Selection-Aware Optimization.}

Table~\ref{tab:rate-cifar-sig} summarizes the Poisoned Selection Rate ($R_{select}$) at the first and last training epochs across different acquisition functions on the CIFAR-10 dataset using the SIG trigger with a poisoning ratio of 0.5\%. The results of other settings can be found in the Appendix.

\begin{table}[tbp]
\small
  \setlength{\tabcolsep}{1.2mm}
  \centering
  \caption{Poisoned Selection Rate (\%) at the First and Last Training Epochs on CIFAR-10 under the SIG Attack (0.5\% Poisoning), across different acquisition functions and optimization iterations. Iter refers to optimization iteration.}
    %\resizebox{\linewidth}{!}{
    \begin{tabular}{cc|cccc}
    \toprule
    Acquisition & Epoch & Iter 0 & Iter 5 & Iter 10 & Iter 15 \\
    \midrule
    \multirow{2}[2]{*}{Random} & 0     & 1.6   & 0.6   & 1.4   & 0.8 \\
          & 9    & 7.6   & 10.0  & 9.4  & 8.6 \\
    \midrule
    \multirow{2}[2]{*}{Entropy} & 0     & 23.8  & 75.4  & 84.6  & 89.4 \\
          & 9    & 29.1  & 75.9  & 84.8  & 89.4 \\
    \midrule
    \multirow{2}[2]{*}{Margin} & 0     & 4.6   & 18.0  & 26.2  & 32.4 \\
          & 9    & 14.6  & 24.3  & 30.1  & 35.5 \\
    \midrule
    \multirow{2}[2]{*}{Least Confidence} & 0     & 23.8  & 78.8  & 89.4  & 95.8 \\
          & 9    & 28.5  & 79.7  & 89.5  & 95.8 \\
    \bottomrule
    \end{tabular}%
  \label{tab:rate-cifar-sig}%
\end{table}%

Comparing the results of iteration 0 (no optimization) with iterations 5, 10, and 15. We observe that \framework substantially increases $R_{select}$ when using uncertainty-based acquisition functions~(e.g., from 29.1\% to 89.4\% under Entropy-based acquisition), while no significant change is observed under Random selection. By computing the correlation between $R_{select}$ and ASR scores~(shown in the Appendix) across training epochs for iterations 5, 10, and 15, we find a positive correlation of 0.698 ($p < 0.001$), indicating a strong relationship between $R_{select}$ and attack success. %This conclusion is further supported by comparing the trends in Table~\ref{tab:rate-cifar-sig} and Figure~\ref{fig:asr-cifar-sig}.

Considering the $R_{select}$ under different training epochs, the results show that the difference between these two epochs is negligible. For instance, across all uncertainty-based acquisition functions and optimization settings, the average change in $R_{select}$ between epoch 0 and epoch 9 is less than 2.9\%. We attribute this to the shift in model decision boundaries after the first epoch of training, which may reduce the uncertainty of poisoned samples that were originally optimized based on the previous model. These findings suggest that, if the attacker can only manipulate samples before injecting them into the AL pool, it is critical to ensure that as many poisoned samples as possible are selected in the very first epoch.

\begin{figure*}[t]
\centering
\includegraphics[width=\textwidth]{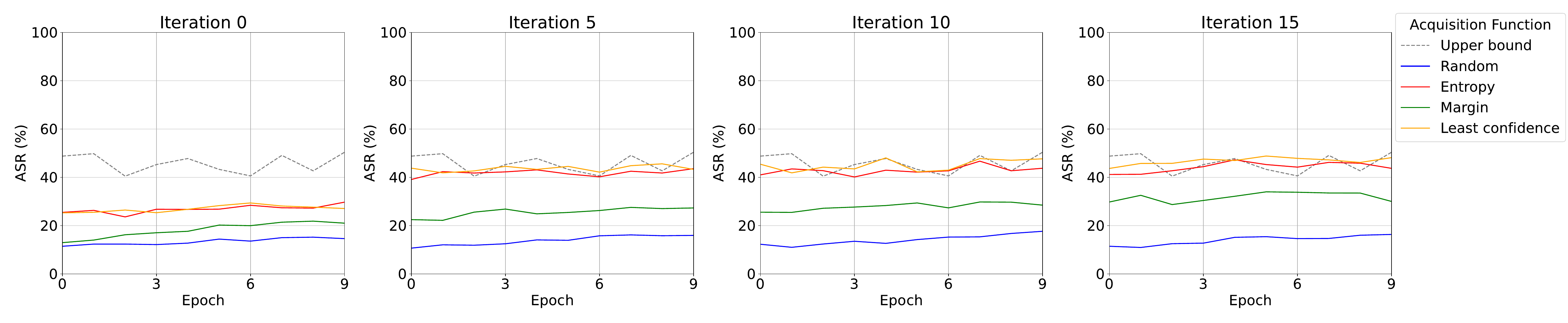} 
\caption{ASR (\%) over training epochs on CIFAR-10 under the SIG trigger with 0.5\% poisoning. Colored lines represent different acquisition functions, while the dashed line denotes the estimated upper bound achieved when all poisoned samples are selected in the first AL epoch.}
\label{fig:asr-cifar-sig}
\end{figure*}

Figure~\ref{fig:asr-cifar-sig} depicts the ASR scores achieved by \framework over different iterations. The dashed line represents a control experiment in which all poisoned samples are manually forced to be selected in the first active learning epoch. In this setting, the poisoned samples participate in training from the beginning of the loop, providing an estimated upper bound for ASR under full exposure. The estimated ASR values are also averaged over attacks targeting all classes. Table~\ref{tab:asr-cifar-sig} reports the final ASR achieved by each acquisition function under different optimization iterations, including the estimated upper bound for reference. %More results are put in the Appendix.

% Colors represent different acquisition functions: red for Entropy, green for Margin, yellow for Least Confidence, and blue for Random.

% Figure~\ref{fig:asr-cifar-sig} shows the ASR over training epochs on the CIFAR-10 dataset using the SIG trigger with a poisoning ratio of 0.5\%. We evaluate \framework under different numbers of optimization iterations: 0 (no optimization), 5, 10, and 15. The x-axis denotes training epochs, and the y-axis represents ASR. The dashed line represents a control experiment in which all poisoned samples are manually forced to be selected in the first active learning round. In this setting, the poisoned samples participate in training from the beginning of the loop, providing an estimated upper bound for ASR under full exposure. The estimated ASR values are also averaged over attacks targeting all classes.
% Colors represent different acquisition functions: red for entropy, green for margin, yellow for least confidence, and blue for random. Table~\ref{tab:asr-cifar-sig} reports the final ASR achieved by each acquisition function under different optimization rounds, including the estimated upper bound for reference.

\begin{table}[t]
\small
  \centering
  \caption{Final ASR (\%) on CIFAR-10 under the SIG attack with 0.5\% poisoning for different acquisition functions and optimization rounds. The first row reports the estimated upper bound, where all poisoned samples are selected in the first AL epoch. Iter refers to optimization iteration.}
    \begin{tabular}{c|cccc}
    \toprule
    ASR (\%) & Iter 0 & Iter 5 & Iter 10 & Iter 15 \\
    \midrule
    Upper bound & 46.0  & 46.0  & 46.0  & 46.0 \\
    Random & 14.6  & 16.0  & 17.7  & 16.4 \\
    Entropy & 29.7  & 43.6  & 43.8  & 43.7 \\
    Margin & 21.0  & 27.3  & 28.5  & 30.0 \\
    Least Confidence & 27.1  & 43.3  & 47.7  & 48.2 \\
    \bottomrule
    \end{tabular}%
    
  \label{tab:asr-cifar-sig}%
\end{table}%

We observe that \framework significantly improves ASR under Entropy and Least Confidence selection, approaching or even surpassing the estimated upper bound. For example, \framework achieves  47.7\% and 48.2\% ASRs on Least Confidence with 10 and 15 optimization iterations, respectively---exceeding the estimated upper bound of 46.0\%. Margin also benefits from optimization. Though its gains are more modest, it still maintains a clear advantage over Random. As expected, Random selection shows no substantial improvement with additional optimization.

% Least Confidence achieves ASRs of 47\% and 58\% with 10 and 15 optimization iterations, respectively---exceeding the estimated upper bound of 46\%. Margin also benefits from optimization. Though its gains are more modest, it still maintains a clear advantage over Random. As expected, Random selection shows no substantial improvement with additional optimization.

%Notably, the ASR differences across different optimization rounds are relatively small, indicating that 5 rounds of optimization are sufficient to achieve stable and effective results. While further rounds offer diminishing returns, we still observe slight ASR gains under margin and least confidence as the number of optimization iterations increases.
Notably, the differences in ASR across different optimization iterations are relatively small, which aligns with our observations from Table~\ref{tab:rate-cifar-sig}: when $R_{select}$ ceases to increase significantly, the ASR also plateaus. This suggests that five iterations of optimization are sufficient to achieve stable and effective results. Nevertheless, although additional optimization iterations yield diminishing returns, we still observe slight ASR improvements under the Margin and Least Confidence acquisition functions, as their corresponding $R_{select}$ values continue to grow slightly.

We further report both $Acc_{ID}$ and $Acc_{OOD}$ results in the Appendix. To investigate whether \framework impacts the model's performance on the original ID data, we plot the $Acc_{ID}$ curve over training epochs under the poisoned setting. The results show that $Acc_{ID}$ remains stable throughout training, with no noticeable degradation and even slight improvements in some cases.  
To assess whether the attack degrades generalization on clean OOD data, we conduct a control experiment where the AL model is trained using the original, unpoisoned OOD samples. The resulting accuracy curve is compared against the poisoned training curve. We observe no significant difference between the two.  
These findings suggest that our clean-label attack does not introduce abnormal accuracy drops or fluctuations in either $Acc_{ID}$ or $Acc_{OOD}$, making it difficult to detect through standard performance monitoring.

\paragraph{Comparison Across Backdoor Attack Methods.}

\begin{figure}[t]
\centering
\includegraphics[width=\columnwidth]{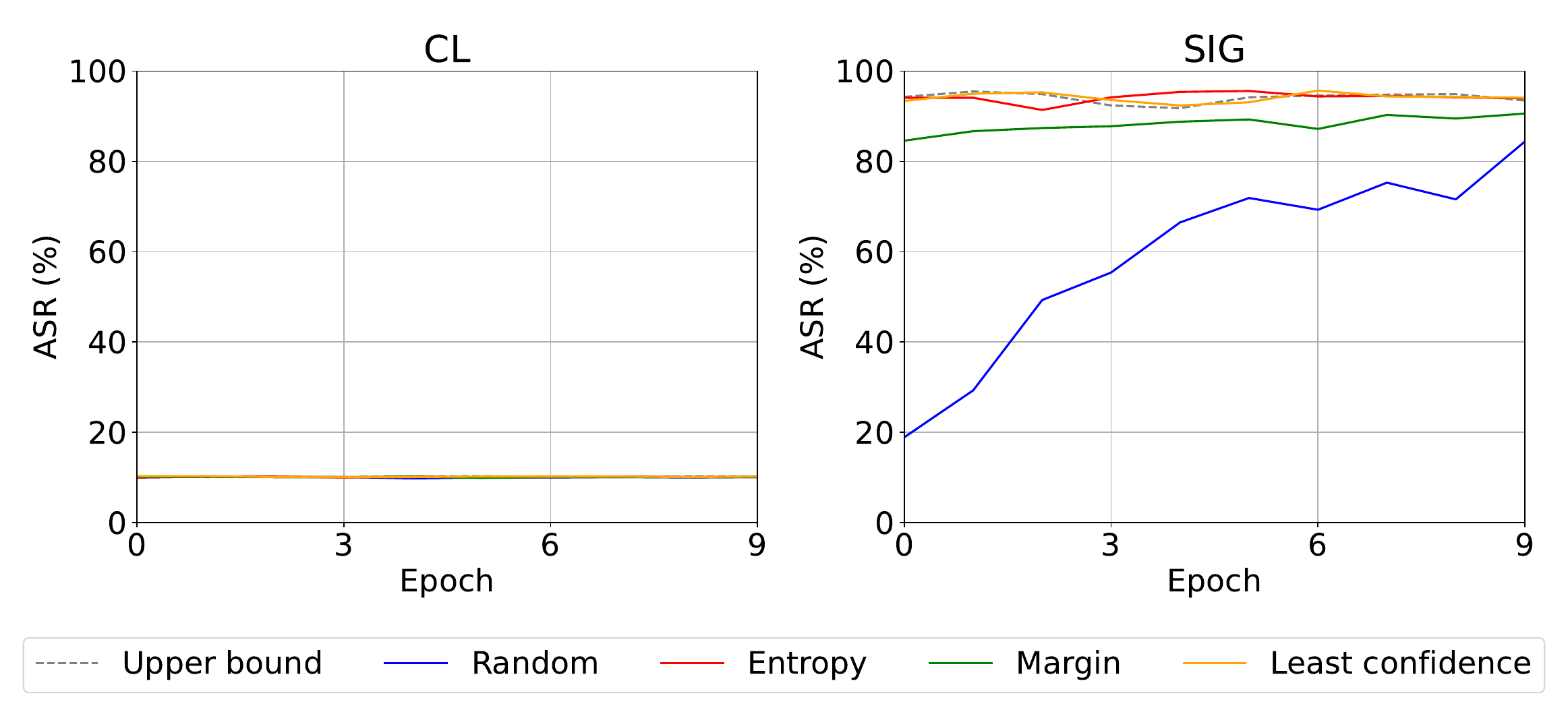} 
\caption{ASR (\%) over training epochs on SVHN under the CL and SIG trigger with 1.0\% poisoning and 15 optimization iterations. Colored lines represent different acquisition functions, while the dashed line denotes the estimated upper bound achieved when all poisoned samples are selected in the first AL epoch.}
\label{fig:svhn-sig-cl}
\end{figure}

We compare two clean-label backdoor attack methods: CL and SIG. While SIG consistently achieves reasonably strong ASR across all datasets, CL proves ineffective as an attack method under the active learning setting. 

Figure~\ref{fig:svhn-sig-cl} illustrates the ASR over training epochs on the CIFAR-10 dataset for both CL and SIG triggers, with a 1\% poisoning ratio and 15 optimization iterations. For reference, we also include the estimated upper bound---obtained by assuming all poisoned samples are selected in the first AL epoch.
The results reveal a stark contrast between the two methods. Under CL, the ASR remains consistently around 10\% throughout training. Even under the estimated upper bound condition, the ASR fails to exceed this level, indicating that CL cannot induce misclassification beyond the poisoned sample’s original class. In contrast, SIG achieves significantly higher ASRs, with attacks optimized by \framework approaching the estimated upper bound, which itself is close to 100\%. More results are put in the Appendix.

These findings demonstrate that in active learning scenarios, SIG is substantially more effective and robust than CL for clean-label backdoor injection---likely due to its stronger and more learnable trigger design.

\begin{figure}[t]
\centering
\includegraphics[width=\columnwidth]{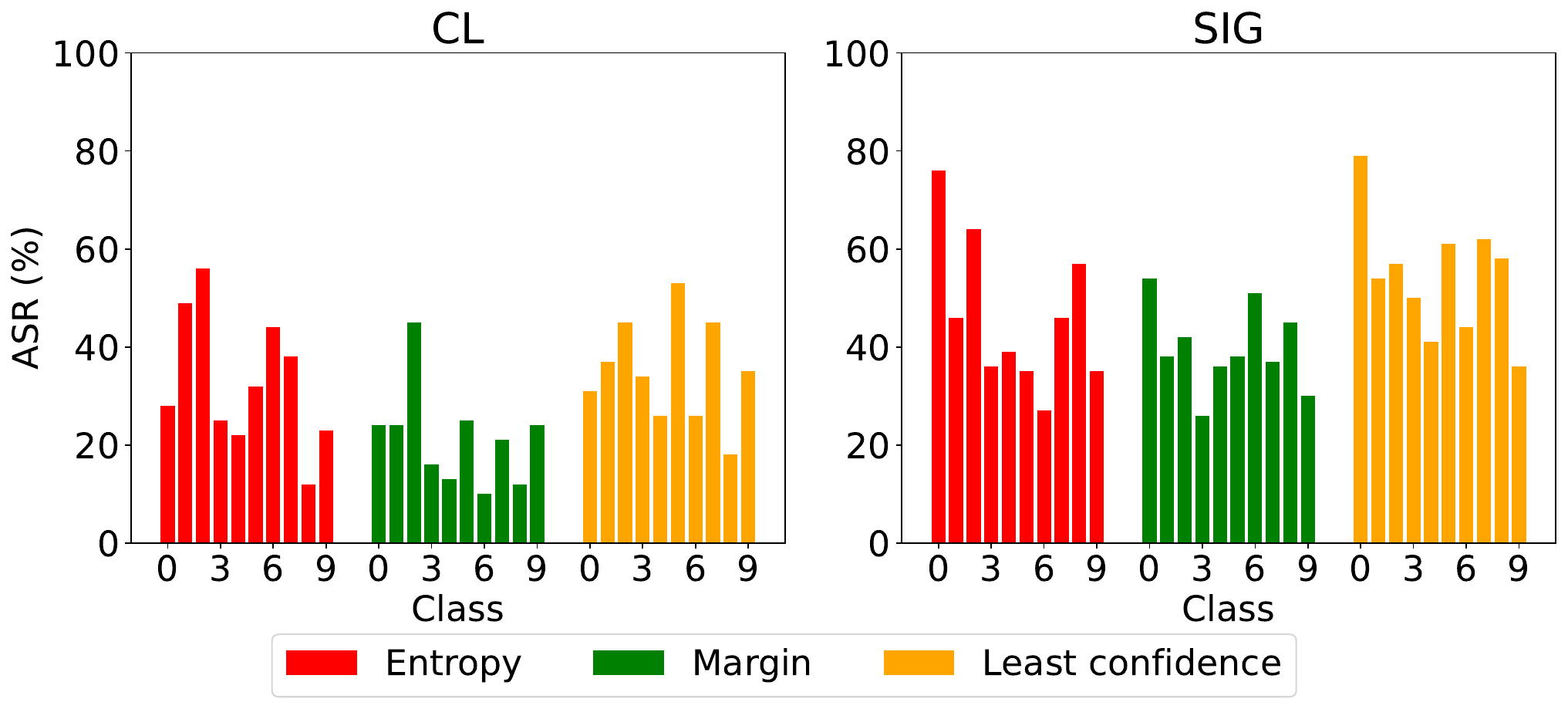} 
\caption{ASR (\%) of epoch 9 over different classes on CIFAR-10 under the CL and SIG trigger with 1.0\% poisoning and 15 optimization iterations. Colored lines represent different acquisition functions, while the dashed line denotes the estimated upper bound achieved when all poisoned samples are selected in the first AL epoch.}
\label{fig:class_study}
\end{figure}

\paragraph{ASR Difference Across Target Classes.}
Figure~\ref{fig:class_study} presents the class-wise ASR on training epoch 9 of CIFAR-10 under the CL and SIG triggers with a 1.0\% poisoning ratio. Each bar corresponds to the ASR for a specific target class when that class is selected as the attack target. 

We observe class-dependent differences in attack difficulty across methods. For example, under all three uncertainty-based acquisition functions, CL consistently yields high ASR on class label 2, while SIG achieves its highest ASR on class label 0, showing consistent behavior within each method. Interestingly, SIG performs well on class label 8, where CL fails to achieve high ASR, further highlighting the differing effectiveness of the two triggers across target classes.

The results reveal a clear contrast in attack generality between the two trigger types. Compared to CL, the SIG trigger achieves consistently higher ASRs across nearly all classes and acquisition functions, demonstrating stronger and more stable attack performance. In particular, under the SIG trigger, most target classes achieve ASRs above 30\%, with more than half exceeding 50\%, and some reaching over 70\%. In contrast, under the CL trigger, ASRs for more than half of the classes fail to surpass 30\%, indicating that SIG generalizes better across different semantic categories. More results can be found in the Appendix.

\section{Discussion}
%selection attack成为新攻击面的影响，不止AL框架下可被攻击
%diversity selection的扩展
%黑盒的可能性/迁移性的讨论
%防御的可能策略
\paragraph{Broader Implications of Selection-Aware Attacks.}
Our work highlights a new and underexplored attack surface: the acquisition function itself.  
While our study focuses on AL, the vulnerability is not confined to AL frameworks.  
Any machine learning pipeline that leverages acquisition function—such as test case selection, data valuation, repair, and retraining strategies—could be similarly exploited.  
This calls for a re-evaluation of the trust placed in selection mechanisms, which have traditionally been seen as purely efficiency-enhancing components.

\paragraph{Extension to Other Acquisition Functions.}
In this work, we focus on uncertainty-based acquisition functions, which are the most widely used category in AL literature.  
However, other acquisition functions—such as diversity-based~\cite{sener2017active, gal2017deep} or hybrid methods~\cite{ash2019deep}—may also be vulnerable.  
Attacking such methods would require new criteria for identifying and optimizing poisoning samples, potentially relying on feature space distances or clustering behavior.  
We leave this as an important direction for future work.

\paragraph{Transferability and Black-box Potential.}
% To explore the black-box applicability of our attack, we evaluated its transferability across models with similar accuracy but different architectures or random initializations.  
% We observed that both the number of selected poisoned samples and the overall attack success rate under uncertainty-based acquisition functions become comparable to Random selection in the transfer setting.  
% This holds even when adopting common transfer techniques such as ensembling, gradient-based approximation, or input-space perturbation diversity.  
% Notably, the samples with high entropy under one model do not consistently exhibit high entropy under another, highlighting a key limitation: uncertainty-based selection is highly sensitive to the decision boundary geometry, which may differ substantially across models even with similar performance.  
% These findings suggest that effective selection-aware attacks in the black-box setting remain a challenging open problem, and further research is needed to design transferable or model-agnostic poisoning techniques.

To explore the black-box applicability of our attack, we evaluated its transferability across models with different architectures or random initializations.  
We observed that both the poisoned selection rate and ASR under uncertainty-based acquisition functions become comparable to Random selection in the transfer setting.  
This holds even when adopting common transfer techniques based on ensembling, gradients, or input transformations. 
Notably, high-entropy samples are not consistent across models, highlighting a key limitation: uncertainty-based selection is highly sensitive to model-specific decision boundaries, even when performance is similar. 
These findings suggest that effective selection-aware attacks in the black-box setting remain a challenging open problem, and further research is needed to design transferable or model-agnostic poisoning techniques. 

\paragraph{Toward Possible Defenses.}
Our findings underscore the need for defenses that explicitly consider vulnerabilities arising from the selection process.  
Potential countermeasures include entropy regularization~\cite{pereyra2017regularizing} to mitigate model overconfidence, and anomaly detection mechanisms to identify suspicious samples selected during AL epochs.  
Another promising direction is to randomize or ensemble multiple acquisition functions, thereby limiting the attacker's ability to optimize against any fixed selection rule.  

Overall, our study issues a new warning for the security of AI systems: acquisition functions, while intended to improve efficiency, can inadvertently introduce exploitable attack surfaces.  
To build truly robust AL systems, it is essential to integrate selection-aware threat modeling into the design of secure learning pipelines.

%\section{Related Work}
%can delete if no space left
%AL security
%backdoor attack

\section{Conclusion}

In this paper, we introduced a new type of attack for deep learning -- poisoning attack for active learning. Leveraging the acquisition function as the attack surface, we proposed a novel framework, \framework, to inject poisoned samples into the training set of active learning models. To increase the attack efficiency, we designed a selection-aware optimization algorithm to maximize the probability of poisoned samples being selected. Evaluation results demonstrated that \framework is effective in attacking active learning, highlighting the need for secure active learning in the future.

% \section{Acknowledgments}

\clearpage

\bibliography{aaai2026}

% \input{ReproducibilityChecklist}

% \cleardoublepage
\clearpage

\section{Appendix}
\label{sec:appendix}

\subsection{Poisoned Selection Rate}
Table 3–14 present the Poisoned Selection Rate at the First and Last Training Epochs across all settings.

\begin{table}[ht]
\small
  \setlength{\tabcolsep}{1.2mm}
  \centering
  \caption{Poisoned Selection Rate (\%) on CIFAR-10 under the CL Attack (0.5\% Poisoning)}
    %\resizebox{\linewidth}{!}{
    \begin{tabular}{cc|cccc}
    \toprule
    Acquisition & Epoch & Iter 0 & Iter 5 & Iter 10 & Iter 15 \\
    \midrule
    \multirow{2}[2]{*}{Random} & 0     & 1.0   & 1.2   & 0.6   & 1.2 \\
          & 9     & 11.2  & 7.2   & 8.0   & 11.6 \\
    \midrule
    \multirow{2}[2]{*}{Entropy} & 0     & 5.8   & 46.0  & 71.4  & 80.8 \\
          & 9     & 14.9  & 47.1  & 71.7  & 81.0 \\
    \midrule
    \multirow{2}[2]{*}{Margin} & 0     & 6.2   & 12.6  & 22.8  & 25.6 \\
          & 9     & 14.2  & 21.5  & 28.4  & 31.3 \\
    \midrule
    \multirow{2}[2]{*}{Least Confidence} & 0     & 7.2   & 49.0  & 72.4  & 88.6 \\
          & 9     & 15.5  & 51.3  & 73.1  & 88.8 \\
    \bottomrule
    \end{tabular}%

  \label{tab:rate-cifar-cl-5}%
\end{table}%

\begin{table}[ht]
\small
  \setlength{\tabcolsep}{1.2mm}
  \centering
  \caption{Poisoned Selection Rate (\%) on Fashion-MNIST under the CL Attack (0.5\% Poisoning)}
    %\resizebox{\linewidth}{!}{
    \begin{tabular}{cc|cccc}
    \toprule
    Acquisition & Epoch & Iter 0 & Iter 5 & Iter 10 & Iter 15 \\
    \midrule
    \multirow{2}[2]{*}{Random} & 0     & 1.8   & 1.4   & 1.2   & 1.6  \\
          & 9     & 12.8  & 8.6   & 10.6  & 9.6  \\
    \midrule
    \multirow{2}[2]{*}{Entropy} & 0     & 0.0   & 18.2  & 40.6  & 60.6  \\
          & 9     & 2.0   & 25.0  & 44.6  & 62.2  \\
    \midrule
    \multirow{2}[2]{*}{Margin} & 0     & 0.0   & 3.4   & 7.2   & 8.2  \\
          & 9     & 6.2   & 11.0  & 14.0  & 13.4  \\
    \midrule
    \multirow{2}[2]{*}{Least Confidence} & 0     & 0.0   & 23.0  & 47.4  & 67.6  \\
          & 9     & 3.2   & 28.0  & 49.6  & 68.4  \\
    \bottomrule
    \end{tabular}%

  \label{tab:rate-fm-cl-5}%
\end{table}%

\begin{table}[ht]
\small
  \setlength{\tabcolsep}{1.2mm}
  \centering
  \caption{Poisoned Selection Rate (\%) on SVHN under the CL Attack (0.5\% Poisoning)}
    %\resizebox{\linewidth}{!}{
    \begin{tabular}{cc|cccc}
    \toprule
    Acquisition & Epoch & Iter 0 & Iter 5 & Iter 10 & Iter 15 \\
    \midrule
    \multirow{2}[2]{*}{Random} & 0     & 1.4   & 0.2   & 0.6   & 1.0  \\
          & 9     & 13.0  & 8.4   & 8.6   & 11.6  \\
    \midrule
    \multirow{2}[2]{*}{Entropy} & 0     & 85.2  & 100.0  & 100.0  & 100.0  \\
          & 9     & 85.4  & 100.0  & 100.0  & 100.0  \\
    \midrule
    \multirow{2}[2]{*}{Margin} & 0     & 10.2  & 14.4  & 19.4  & 25.4  \\
          & 9     & 18.2  & 21.2  & 24.8  & 31.0  \\
    \midrule
    \multirow{2}[2]{*}{Least Confidence} & 0     & 61.0  & 95.0  & 98.4  & 99.8  \\
          & 9     & 61.8  & 95.0  & 98.4  & 99.8  \\
    \bottomrule
    \end{tabular}%

  \label{tab:rate-svhn-cl-5}%
\end{table}%

\newpage
\begin{table}[ht]
\small
  \setlength{\tabcolsep}{1.2mm}
  \centering
  \caption{Poisoned Selection Rate (\%) on CIFAR-10 under the SIG Attack (0.5\% Poisoning)}
    %\resizebox{\linewidth}{!}{
    \begin{tabular}{cc|cccc}
    \toprule
    Acquisition & Epoch & Iter 0 & Iter 5 & Iter 10 & Iter 15 \\
    \midrule
    \multirow{2}[2]{*}{Random} & 0     & 1.6   & 0.6   & 1.4   & 0.8 \\
          & 9    & 7.6   & 10.0  & 9.4  & 8.6 \\
    \midrule
    \multirow{2}[2]{*}{Entropy} & 0     & 23.8  & 75.4  & 84.6  & 89.4 \\
          & 9    & 29.1  & 75.9  & 84.8  & 89.4 \\
    \midrule
    \multirow{2}[2]{*}{Margin} & 0     & 4.6   & 18.0  & 26.2  & 32.4 \\
          & 9    & 14.6  & 24.3  & 30.1  & 35.5 \\
    \midrule
    \multirow{2}[2]{*}{Least Confidence} & 0     & 23.8  & 78.8  & 89.4  & 95.8 \\
          & 9    & 28.5  & 79.7  & 89.5  & 95.8 \\
    \bottomrule
    \end{tabular}%
  \label{tab:rate-cifar-sig-5}%
\end{table}%

\begin{table}[ht]
\small
  \setlength{\tabcolsep}{1.2mm}
  \centering
  \caption{Poisoned Selection Rate (\%) on Fashion-MNIST under the SIG Attack (0.5\% Poisoning)}
    %\resizebox{\linewidth}{!}{
    \begin{tabular}{cc|cccc}
    \toprule
    Acquisition & Epoch & Iter 0 & Iter 5 & Iter 10 & Iter 15 \\
    \midrule
    \multirow{2}[2]{*}{Random} & 0     & 1.4   & 1.8   & 0.4   & 0.2  \\
          & 9     & 9.6   & 12.4  & 7.0   & 9.4  \\
    \midrule
    \multirow{2}[2]{*}{Entropy} & 0     & 15.2  & 53.2  & 72.0  & 86.0  \\
          & 9     & 20.6  & 54.2  & 72.2  & 86.2  \\
    \midrule
    \multirow{2}[2]{*}{Margin} & 0     & 2.6   & 4.6   & 5.2   & 7.2  \\
          & 9     & 8.4   & 11.4  & 10.8  & 11.8  \\
    \midrule
    \multirow{2}[2]{*}{Least Confidence} & 0     & 9.6   & 47.6  & 69.8  & 85.4  \\
          & 9     & 16.2  & 48.4  & 70.4  & 85.4  \\
    \bottomrule
    \end{tabular}%
  \label{tab:rate-fm-sig-5}%
\end{table}%

\begin{table}[ht]
\small
  \setlength{\tabcolsep}{1.2mm}
  \centering
  \caption{Poisoned Selection Rate (\%) on SVHN under the SIG Attack (0.5\% Poisoning)}
    %\resizebox{\linewidth}{!}{
    \begin{tabular}{cc|cccc}
    \toprule
    Acquisition & Epoch & Iter 0 & Iter 5 & Iter 10 & Iter 15 \\
    \midrule
    \multirow{2}[2]{*}{Random} & 0     & 1.4   & 0.2   & 0.6   & 1.0  \\
          & 9     & 13.0  & 8.4   & 8.6   & 11.6  \\
    \midrule
    \multirow{2}[2]{*}{Entropy} & 0     & 85.2  & 100.0  & 100.0  & 100.0  \\
          & 9     & 85.4  & 100.0  & 100.0  & 100.0  \\
    \midrule
    \multirow{2}[2]{*}{Margin} & 0     & 10.2  & 14.4  & 19.4  & 25.4  \\
          & 9     & 18.2  & 21.2  & 24.8  & 31.0  \\
    \midrule
    \multirow{2}[2]{*}{Least Confidence} & 0     & 61.0  & 95.0  & 98.4  & 99.8  \\
          & 9     & 61.8  & 95.0  & 98.4  & 99.8  \\
    \bottomrule
    \end{tabular}%
  % \label{tab:rate-cifar-sig}%
\end{table}%

\clearpage
\begin{table}[ht]
\small
  \setlength{\tabcolsep}{1.2mm}
  \centering
  \caption{Poisoned Selection Rate (\%) on CIFAR-10 under the CL Attack (1.0\% Poisoning)}
    %\resizebox{\linewidth}{!}{
    \begin{tabular}{cc|cccc}
    \toprule
    Acquisition & Epoch & Iter 0 & Iter 5 & Iter 10 & Iter 15 \\
    \midrule
    \multirow{2}[2]{*}{Random} & 0     & 1.2   & 1.0   & 1.0   & 0.5  \\
          & 9     & 9.5   & 11.0  & 10.8  & 8.8  \\
    \midrule
    \multirow{2}[2]{*}{Entropy} & 0     & 3.0   & 45.6  & 65.1  & 73.9  \\
          & 9     & 11.5  & 46.0  & 65.4  & 74.0  \\
    \midrule
    \multirow{2}[2]{*}{Margin} & 0     & 3.7   & 14.3  & 19.7  & 23.9  \\
          & 9     & 10.9  & 19.4  & 23.3  & 26.6  \\
    \midrule
    \multirow{2}[2]{*}{Least Confidence} & 0     & 3.6   & 44.0  & 67.9  & 77.4  \\
          & 9     & 11.3  & 44.6  & 68.0  & 77.5  \\
    \bottomrule
    \end{tabular}%

  % \label{tab:rate-cifar-sig}%
\end{table}%

\begin{table}[ht]
\small
  \setlength{\tabcolsep}{1.2mm}
  \centering
  \caption{Poisoned Selection Rate (\%) on Fashion-MNIST under the CL Attack (1.0\% Poisoning)}
    %\resizebox{\linewidth}{!}{
    \begin{tabular}{cc|cccc}
    \toprule
    Acquisition & Epoch & Iter 0 & Iter 5 & Iter 10 & Iter 15 \\
    \midrule
    \multirow{2}[2]{*}{Random} & 0     & 1.0   & 1.1   & 0.9   & 0.5  \\
          & 9     & 11.0  & 10.6  & 10.7  & 8.4  \\
    \midrule
    \multirow{2}[2]{*}{Entropy} & 0     & 0.0   & 19.4  & 42.3  & 61.2  \\
          & 9     & 3.0   & 23.6  & 43.8  & 61.8  \\
    \midrule
    \multirow{2}[2]{*}{Margin} & 0     & 0.0   & 4.1   & 6.0   & 8.4  \\
          & 9     & 4.3   & 10.3  & 12.3  & 13.9  \\
    \midrule
    \multirow{2}[2]{*}{Least Confidence} & 0     & 0.0   & 24.8  & 47.9  & 64.5  \\
          & 9     & 3.1   & 26.6  & 48.9  & 64.9  \\
    \bottomrule
    \end{tabular}%

  % \label{tab:rate-cifar-sig}%
\end{table}%

\begin{table}[ht]
\small
  \setlength{\tabcolsep}{1.2mm}
  \centering
  \caption{Poisoned Selection Rate (\%) on SVHN under the CL Attack (1.0\% Poisoning)}
    %\resizebox{\linewidth}{!}{
    \begin{tabular}{cc|cccc}
    \toprule
    Acquisition & Epoch & Iter 0 & Iter 5 & Iter 10 & Iter 15 \\
    \midrule
    \multirow{2}[2]{*}{Random} & 0     & 0.9   & 0.7   & 0.8   & 0.8  \\
          & 9     & 10.3  & 8.4   & 10.0  & 11.1  \\
    \midrule
    \multirow{2}[2]{*}{Entropy} & 0     & 0.0   & 31.5  & 51.2  & 62.3  \\
          & 9     & 10.7  & 42.5  & 57.4  & 67.3  \\
    \midrule
    \multirow{2}[2]{*}{Margin} & 0     & 0.0   & 7.6   & 10.0  & 12.3  \\
          & 9     & 8.3   & 20.6  & 22.7  & 25.6  \\
    \midrule
    \multirow{2}[2]{*}{Least Confidence} & 0     & 0.0   & 33.1  & 50.9  & 64.9  \\
          & 9     & 10.4  & 41.8  & 57.9  & 69.3  \\
    \bottomrule
    \end{tabular}%

  % \label{tab:rate-cifar-sig}%
\end{table}%

\newpage
\begin{table}[ht]
\small
  \setlength{\tabcolsep}{1.2mm}
  \centering
  \caption{Poisoned Selection Rate (\%) on CIFAR-10 under the SIG Attack (1.0\% Poisoning)}
    %\resizebox{\linewidth}{!}{
    \begin{tabular}{cc|cccc}
    \toprule
    Acquisition & Epoch & Iter 0 & Iter 5 & Iter 10 & Iter 15 \\
    \midrule
    \multirow{2}[2]{*}{Random} & 0     & 1.2   & 1.0   & 1.4   & 0.8  \\
          & 9     & 9.3   & 8.4   & 10.1  & 9.4  \\
    \midrule
    \multirow{2}[2]{*}{Entropy} & 0     & 11.9  & 58.2  & 69.4  & 73.4  \\
          & 9     & 16.1  & 58.4  & 69.6  & 73.6  \\
    \midrule
    \multirow{2}[2]{*}{Margin} & 0     & 4.5   & 18.3  & 23.2  & 24.7  \\
          & 9     & 12.1  & 21.7  & 25.9  & 27.0  \\
    \midrule
    \multirow{2}[2]{*}{Least Confidence} & 0     & 12.6  & 62.7  & 76.0  & 81.9  \\
          & 9     & 16.7  & 63.2  & 76.1  & 82.0  \\
    \bottomrule
    \end{tabular}%

  % \label{tab:rate-cifar-sig}%
\end{table}%

\begin{table}[ht]
\small
  \setlength{\tabcolsep}{1.2mm}
  \centering
  \caption{Poisoned Selection Rate (\%) on Fashion-MNIST under the SIG Attack (1.0\% Poisoning)}
    %\resizebox{\linewidth}{!}{
    \begin{tabular}{cc|cccc}
    \toprule
    Acquisition & Epoch & Iter 0 & Iter 5 & Iter 10 & Iter 15 \\
    \midrule
    \multirow{2}[2]{*}{Random} & 0     & 1.5   & 1.4   & 1.3   & 1.0  \\
          & 9     & 9.1   & 10.3  & 9.5   & 10.1  \\
    \midrule
    \multirow{2}[2]{*}{Entropy} & 0     & 7.8   & 44.6  & 65.7  & 79.4  \\
          & 9     & 12.7  & 45.3  & 65.9  & 79.7  \\
    \midrule
    \multirow{2}[2]{*}{Margin} & 0     & 2.0   & 5.5   & 6.6   & 9.8  \\
          & 9     & 8.2   & 7.5   & 9.8   & 11.5  \\
    \midrule
    \multirow{2}[2]{*}{Least Confidence} & 0     & 5.4   & 41.6  & 61.1  & 76.0  \\
          & 9     & 10.1  & 41.7  & 61.2  & 76.0  \\
    \bottomrule
    \end{tabular}%
  % \label{tab:rate-cifar-sig}%
\end{table}%

\begin{table}[ht]
\small
  \setlength{\tabcolsep}{1.2mm}
  \centering
  \caption{Poisoned Selection Rate (\%) on SVHN under the SIG Attack (1.0\% Poisoning)}
    %\resizebox{\linewidth}{!}{
    \begin{tabular}{cc|cccc}
    \toprule
    Acquisition & Epoch & Iter 0 & Iter 5 & Iter 10 & Iter 15 \\
    \midrule
    \multirow{2}[2]{*}{Random} & 0     & 0.6   & 1.6   & 0.9   & 0.9  \\
          & 9     & 10.4  & 9.3   & 9.0   & 11.0  \\
    \midrule
    \multirow{2}[2]{*}{Entropy} & 0     & 43.6  & 74.8  & 82.5  & 86.1  \\
          & 9     & 44.1  & 75.0  & 82.7  & 86.2  \\
    \midrule
    \multirow{2}[2]{*}{Margin} & 0     & 7.2   & 14.7  & 17.0  & 21.1  \\
          & 9     & 13.3  & 17.7  & 19.6  & 23.6  \\
    \midrule
    \multirow{2}[2]{*}{Least Confidence} & 0     & 37.1  & 74.9  & 84.3  & 88.4  \\
          & 9     & 38.5  & 75.1  & 84.3  & 88.6  \\
    \bottomrule
    \end{tabular}%
  % \label{tab:rate-cifar-sig}%
\end{table}%

\clearpage
\subsection{Attack success rate}
Table 15–26 present the Final ASR (\%) for different acquisition functions and optimization rounds. Figure 5-16 show the ASR (\%) across epochs. Table 39 shows the Poisoned Selection Rate and the corresponding ASR on the CIFAR-10 dataset using the SIG trigger with a poisoning ratio of 0.5\%.

\begin{table}[ht]
\small
  \centering
  \caption{Final ASR (\%) on CIFAR-10 under the CL attack with 0.5\% poisoning for different acquisition functions and optimization rounds. The first row reports the estimated upper bound, where all poisoned samples are selected in the first AL epoch. Iter refers to optimization iteration.}
   
    \begin{tabular}{c|cccc}
    \toprule
    ASR (\%) & Iter 0 & Iter 5 & Iter 10 & Iter 15 \\
    \midrule
    Upper bound & 55.1  & 55.1  & 55.1  & 55.1  \\
    Random & 11.6  & 10.4  & 10.4  & 10.8  \\
    Entropy & 12.0  & 16.7  & 20.9  & 25.0  \\
    Margin & 13.1  & 11.9  & 12.5  & 13.7  \\
    Least Confidence & 11.9  & 16.2  & 22.0  & 26.5  \\
    \bottomrule
    \end{tabular}%
    
  % \label{tab:asr-cifar-sig}%
\end{table}%

\begin{table}[ht]
\small
  \centering
  \caption{Final ASR (\%) on Fashion-MNIST under the CL attack with 0.5\% poisoning for different acquisition functions and optimization rounds. The first row reports the estimated upper bound, where all poisoned samples are selected in the first AL epoch. Iter refers to optimization iteration.}
    \begin{tabular}{c|cccc}
    \toprule
    ASR (\%) & Iter 0 & Iter 5 & Iter 10 & Iter 15 \\
    \midrule
    Upper bound & 35.8  & 35.8  & 35.8  & 35.8  \\
    Random & 16.9  & 12.5  & 11.0  & 13.3  \\
    Entropy & 11.3  & 13.0  & 15.7  & 20.2  \\
    Margin & 11.9  & 12.7  & 14.2  & 13.7  \\
    Least Confidence & 10.5  & 14.7  & 17.4  & 24.4  \\
    \bottomrule
    \end{tabular}%

  % \label{tab:asr-cifar-sig}%
\end{table}%

\begin{table}[ht]
\small
  \centering
  \caption{Final ASR (\%) on SVHN under the CL attack with 0.5\% poisoning for different acquisition functions and optimization rounds. The first row reports the estimated upper bound, where all poisoned samples are selected in the first AL epoch. Iter refers to optimization iteration.}
    \begin{tabular}{c|cccc}
    \toprule
    ASR (\%) & Iter 0 & Iter 5 & Iter 10 & Iter 15 \\
    \midrule
    Upper bound & 10.1  & 10.1  & 10.1  & 10.1  \\
    Random & 10.1  & 10.2  & 10.0  & 10.0  \\
    Entropy & 10.1  & 10.0  & 10.1  & 10.0  \\
    Margin & 10.1  & 10.1  & 10.0  & 10.2  \\
    Least Confidence & 10.1  & 10.1  & 10.1  & 10.1  \\
    \bottomrule
    \end{tabular}%
    
  % \label{tab:asr-cifar-sig}%
\end{table}%

\newpage

\begin{table}[ht]
\small
  \centering
  \caption{Final ASR (\%) on CIFAR-10 under the SIG attack with 0.5\% poisoning for different acquisition functions and optimization rounds. The first row reports the estimated upper bound, where all poisoned samples are selected in the first AL epoch. Iter refers to optimization iteration.}
    \begin{tabular}{c|cccc}
    \toprule
    ASR (\%) & Iter 0 & Iter 5 & Iter 10 & Iter 15 \\
    \midrule
    Upper bound & 46.0  & 46.0  & 46.0  & 46.0  \\
    Random & 14.6  & 16.0  & 17.7  & 16.4  \\
    Entropy & 29.7  & 43.6  & 43.8  & 43.7  \\
    Margin & 21.0  & 27.3  & 28.5  & 30.0  \\
    Least Confidence & 27.1  & 43.3  & 47.7  & 48.2  \\
    \bottomrule
    \end{tabular}%
    
  % \label{tab:asr-cifar-sig}%
\end{table}%

\begin{table}[ht]
\small
  \centering
  \caption{Final ASR (\%) on Fashion-MNIST under the SIG attack with 0.5\% poisoning for different acquisition functions and optimization rounds. The first row reports the estimated upper bound, where all poisoned samples are selected in the first AL epoch. Iter refers to optimization iteration.}
    \begin{tabular}{c|cccc}
    \toprule
    ASR (\%) & Iter 0 & Iter 5 & Iter 10 & Iter 15 \\
    \midrule
    Upper bound & 33.3  & 33.3  & 33.3  & 33.3  \\
    Random & 17.0  & 19.3  & 14.9  & 18.1  \\
    Entropy & 20.5  & 25.9  & 27.8  & 29.0  \\
    Margin & 17.5  & 18.5  & 19.4  & 21.7  \\
    Least Confidence & 21.3  & 27.8  & 27.2  & 33.6  \\
    \bottomrule
    \end{tabular}%
    
  % \label{tab:asr-cifar-sig}%
\end{table}%

\begin{table}[ht]
\small
  \centering
  \caption{Final ASR (\%) on SVHN under the SIG attack with 0.5\% poisoning for different acquisition functions and optimization rounds. The first row reports the estimated upper bound, where all poisoned samples are selected in the first AL epoch. Iter refers to optimization iteration.}
    \begin{tabular}{c|cccc}
    \toprule
    ASR (\%) & Iter 0 & Iter 5 & Iter 10 & Iter 15 \\
    \midrule
    Upper bound & 91.6  & 91.6  & 91.6  & 91.6  \\
    Random & 74.0  & 51.1  & 65.7  & 64.6  \\
    Entropy & 90.5  & 90.9  & 93.2  & 93.0  \\
    Margin & 80.9  & 83.8  & 81.5  & 77.3  \\
    Least Confidence & 87.6  & 92.6  & 92.5  & 87.8  \\
    \bottomrule
    \end{tabular}%
    
  % \label{tab:asr-cifar-sig}%
\end{table}%

\newpage

\begin{table}[ht]
\small
  \centering
  \caption{Final ASR (\%) on CIFAR-10 under the CL attack with 1.0\% poisoning for different acquisition functions and optimization rounds. The first row reports the estimated upper bound, where all poisoned samples are selected in the first AL epoch. Iter refers to optimization iteration.}
    
    \begin{tabular}{c|cccc}
    \toprule
    ASR (\%) & Iter 0 & Iter 5 & Iter 10 & Iter 15 \\
    \midrule
    Upper bound & 66.6  & 66.6  & 66.6  & 66.6  \\
    Random & 12.3  & 12.2  & 11.6  & 11.5  \\
    Entropy & 15.1  & 27.4  & 33.3  & 33.8  \\
    Margin & 16.3  & 19.2  & 17.6  & 20.5  \\
    Least Confidence & 13.5  & 27.5  & 33.2  & 34.9  \\
    \bottomrule
    \end{tabular}%
    
  % \label{tab:asr-cifar-sig}%
\end{table}%

\begin{table}[ht]
\small
  \centering
  \caption{Final ASR (\%) on Fashion-MNIST under the CL attack with 1.0\% poisoning for different acquisition functions and optimization rounds. The first row reports the estimated upper bound, where all poisoned samples are selected in the first AL epoch. Iter refers to optimization iteration.}
   \begin{tabular}{c|cccc}
    \toprule
    ASR (\%) & Iter 0 & Iter 5 & Iter 10 & Iter 15 \\
    \midrule
    Upper bound & 47.6  & 47.6  & 47.6  & 47.6  \\
    Random & 17.4  & 13.4  & 13.3  & 13.3  \\
    Entropy & 10.7  & 16.4  & 20.3  & 22.4  \\
    Margin & 14.8  & 15.1  & 15.3  & 15.0  \\
    Least Confidence & 11.5  & 17.0  & 20.4  & 23.8  \\
    \bottomrule
    \end{tabular}%
    
  % \label{tab:asr-cifar-sig}%
\end{table}%

\begin{table}[ht]
\small
  \centering
  \caption{Final ASR (\%) on SVHN under the CL attack with 1.0\% poisoning for different acquisition functions and optimization rounds. The first row reports the estimated upper bound, where all poisoned samples are selected in the first AL epoch. Iter refers to optimization iteration.}
   \begin{tabular}{c|cccc}
    \toprule
    ASR (\%) & Iter 0 & Iter 5 & Iter 10 & Iter 15 \\
    \midrule
    Upper bound & 10.1  & 10.1  & 10.1  & 10.1  \\
    Random & 10.1  & 10.2  & 10.0  & 10.1  \\
    Entropy & 10.0  & 10.1  & 9.9   & 10.1  \\
    Margin & 10.1  & 10.0  & 10.1  & 10.1  \\
    Least Confidence & 10.1  & 10.2  & 10.1  & 10.2  \\
    \bottomrule
    \end{tabular}%
    
  % \label{tab:asr-cifar-sig}%
\end{table}%

\newpage
\begin{table}[ht]
\small
  \centering
  \caption{Final ASR (\%) on CIFAR-10 under the SIG attack with 1.0\% poisoning for different acquisition functions and optimization rounds. The first row reports the estimated upper bound, where all poisoned samples are selected in the first AL epoch. Iter refers to optimization iteration.}
    \begin{tabular}{c|cccc}
    \toprule
    ASR (\%) & Iter 0 & Iter 5 & Iter 10 & Iter 15 \\
    \midrule
    Upper bound & 60.3  & 60.3  & 60.3  & 60.3  \\
    Random & 24.1  & 20.8  & 23.8  & 19.9  \\
    Entropy & 30.5  & 50.3  & 48.3  & 48.1  \\
    Margin & 30.2  & 36.6  & 38.3  & 38.1  \\
    Least Confidence & 34.4  & 50.6  & 53.5  & 52.2  \\
    \bottomrule
    \end{tabular}%

  % \label{tab:asr-cifar-sig}%
\end{table}%

\begin{table}[ht]
\small
  \centering
  \caption{Final ASR (\%) on Fashion-MNIST under the SIG attack with 1.0\% poisoning for different acquisition functions and optimization rounds. The first row reports the estimated upper bound, where all poisoned samples are selected in the first AL epoch. Iter refers to optimization iteration.}
  
    \begin{tabular}{c|cccc}
    \toprule
    ASR (\%) & Iter 0 & Iter 5 & Iter 10 & Iter 15 \\
    \midrule
    Upper bound & 31.5  & 31.5  & 31.5  & 31.5  \\
    Random & 19.5  & 24.0  & 23.1  & 21.6  \\
    Entropy & 24.0  & 25.5  & 31.9  & 34.2  \\
    Margin & 21.3  & 24.9  & 25.8  & 27.0  \\
    Least Confidence & 20.9  & 33.5  & 30.4  & 34.8  \\
    \bottomrule
    \end{tabular}%
    
  % \label{tab:asr-cifar-sig}%
\end{table}%

\begin{table}[ht]
\small
  \centering
  \caption{Final ASR (\%) on SVHN under the SIG attack with 1.0\% poisoning for different acquisition functions and optimization rounds. The first row reports the estimated upper bound, where all poisoned samples are selected in the first AL epoch. Iter refers to optimization iteration.}
   \begin{tabular}{c|cccc}
    \toprule
    ASR (\%) & Iter 0 & Iter 5 & Iter 10 & Iter 15 \\
    \midrule
    Upper bound & 94.3  & 94.3  & 94.3  & 94.3  \\
    Random & 77.1  & 74.5  & 80.0  & 84.4  \\
    Entropy & 86.9  & 93.3  & 91.1  & 94.0  \\
    Margin & 88.0  & 88.8  & 85.7  & 90.6  \\
    Least Confidence & 91.6  & 94.0  & 93.3  & 94.1  \\
    \bottomrule
    \end{tabular}%
    
  % \label{tab:asr-cifar-sig}%
\end{table}%

\begin{figure*}
\centering
\includegraphics[width=\textwidth]{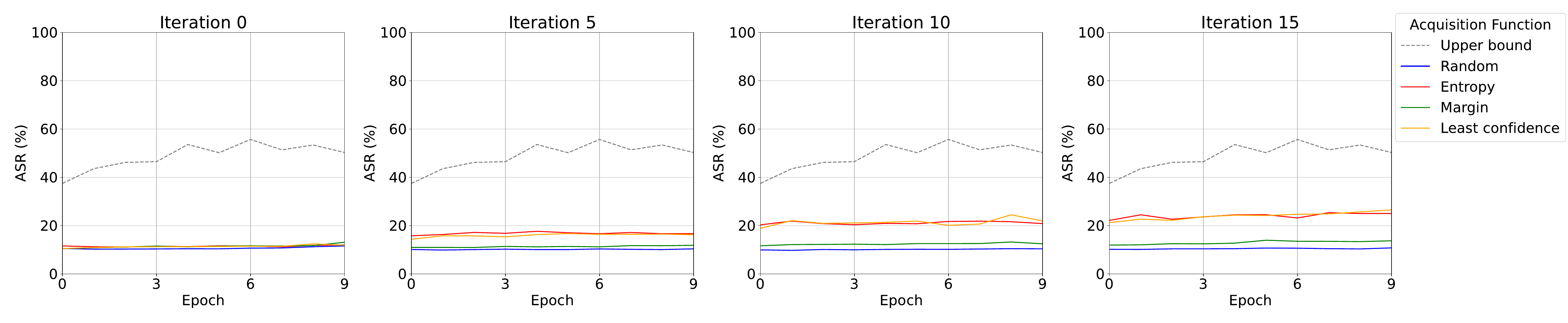} 
\caption{ASR (\%) over training epochs on CIFAR-10 under the CL trigger with 0.5\% poisoning. Colored lines represent different acquisition functions, while the dashed line denotes the estimated upper bound achieved when all poisoned samples are selected in the first AL epoch.}
% \label{fig:asr-cifar-sig}
\end{figure*}

\begin{figure*}
\centering
\includegraphics[width=\textwidth]{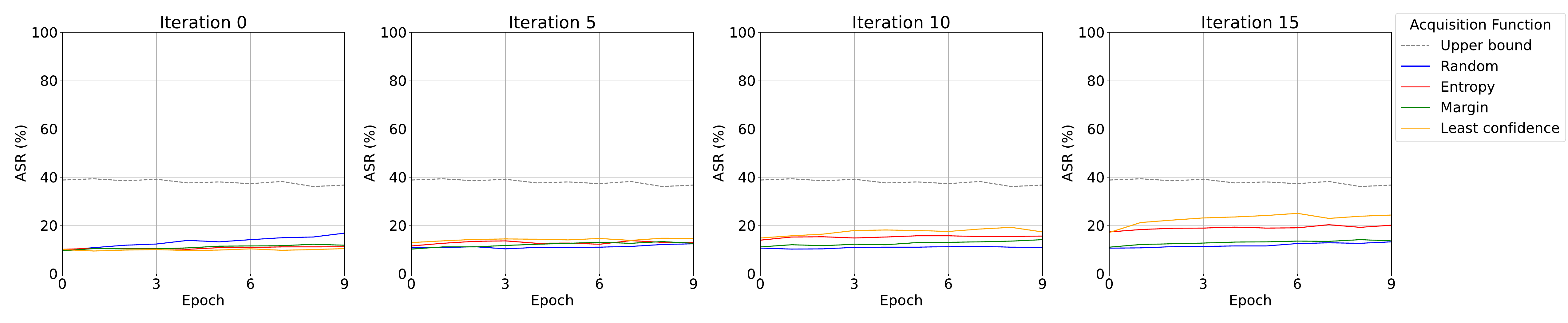} 
\caption{ASR (\%) over training epochs on Fashion-MNIST under the CL trigger with 0.5\% poisoning.}
% \label{fig:asr-cifar-sig}
\end{figure*}

\begin{figure*}
\centering
\includegraphics[width=\textwidth]{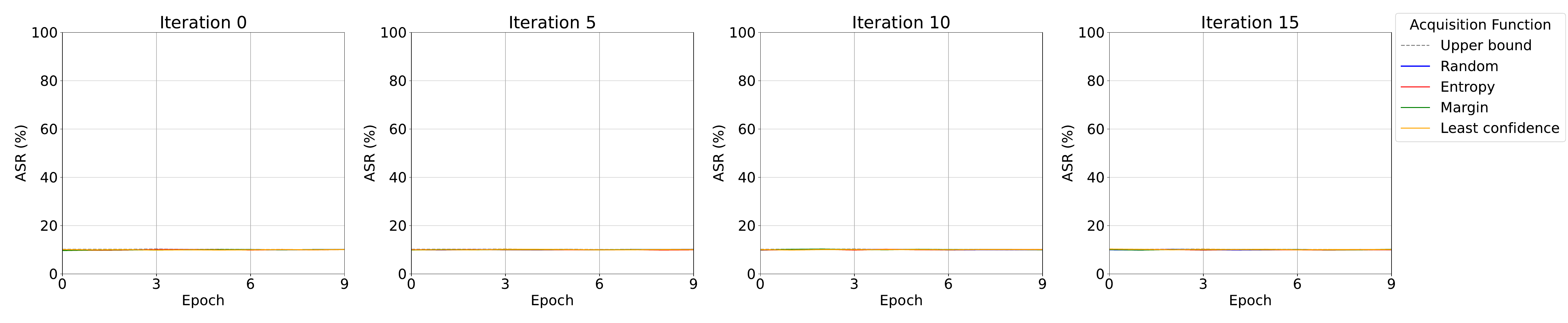} 
\caption{ASR (\%) over training epochs on SVHN under the CL trigger with 0.5\% poisoning.}
% \label{fig:asr-cifar-sig}
\end{figure*}

\begin{figure*}
\centering
\includegraphics[width=\textwidth]{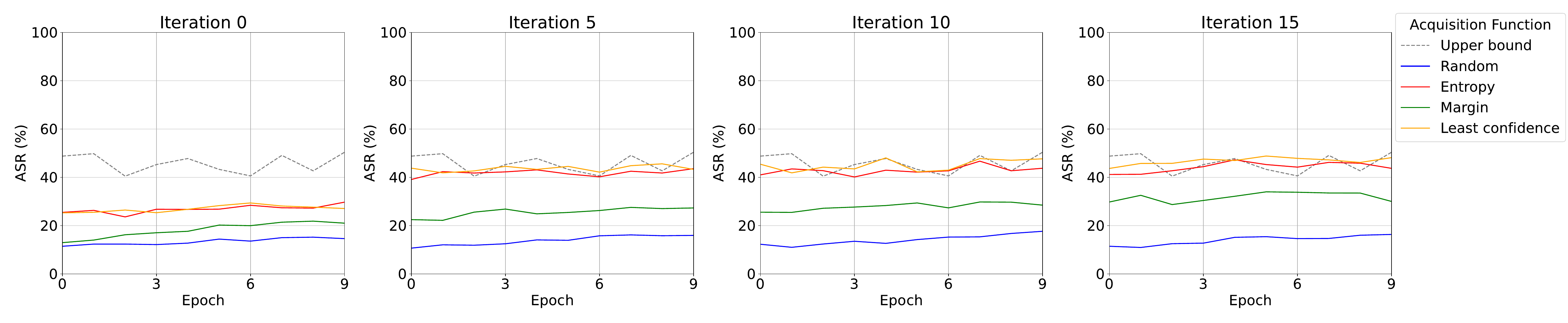} 
\caption{ASR (\%) over training epochs on CIFAR-10 under the SIG trigger with 0.5\% poisoning.}
% \label{fig:asr-cifar-sig}
\end{figure*}

\begin{figure*}
\centering
\includegraphics[width=\textwidth]{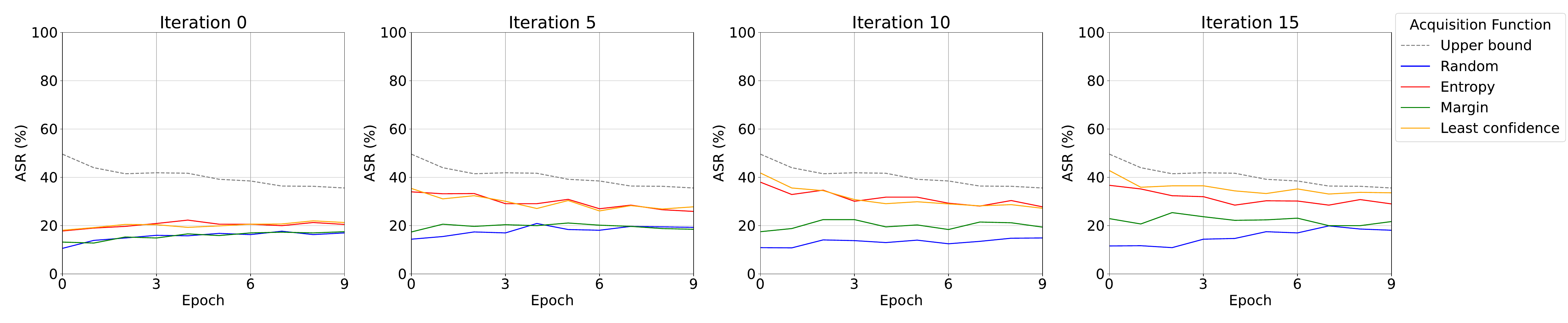} 
\caption{ASR (\%) over training epochs on Fashion-MNIST under the SIG trigger with 0.5\% poisoning.}
% \label{fig:asr-cifar-sig}
\end{figure*}

\begin{figure*}
\centering
\includegraphics[width=\textwidth]{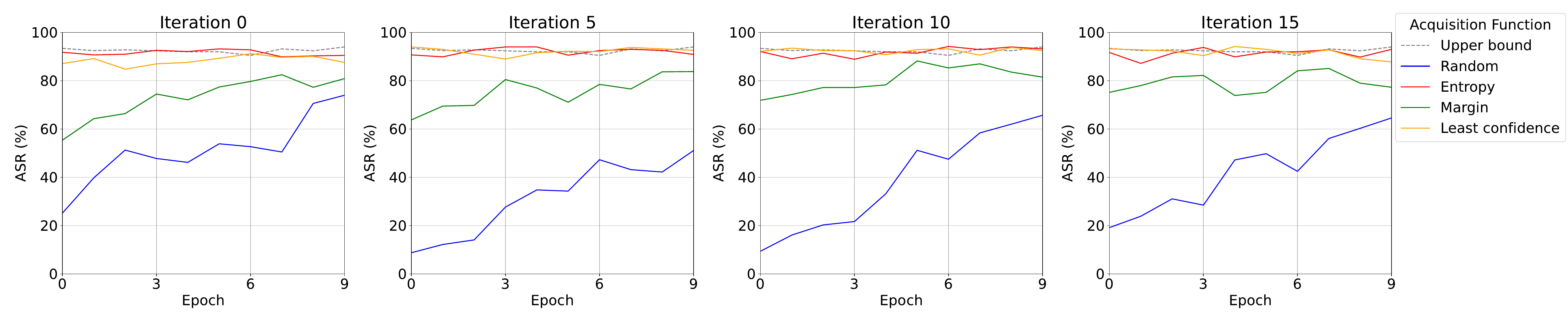} 
\caption{ASR (\%) over training epochs on SVHN under the SIG trigger with 0.5\% poisoning.}
% \label{fig:asr-cifar-sig}
\end{figure*}

\begin{figure*}
\centering
\includegraphics[width=\textwidth]{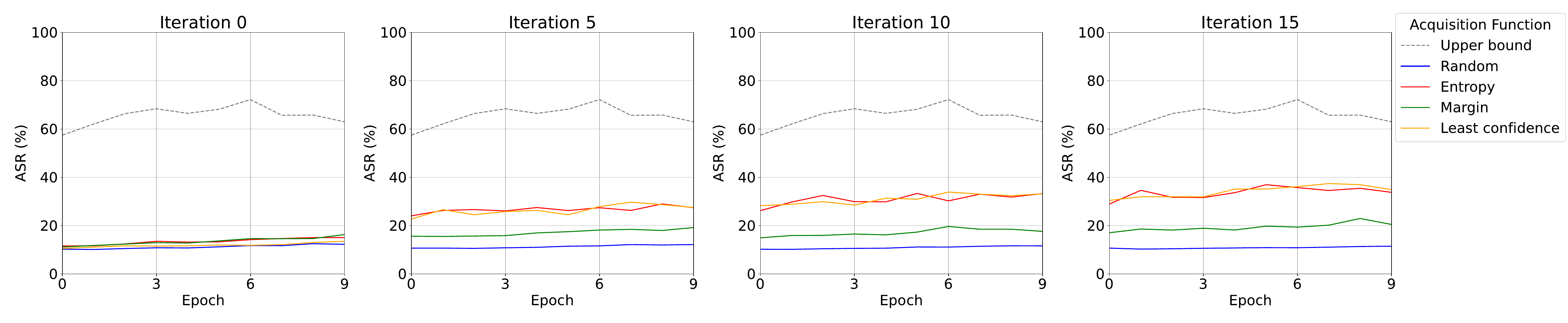} 
\caption{ASR (\%) over training epochs on CIFAR-10 under the CL trigger with 1.0\% poisoning. }
% \label{fig:asr-cifar-sig}
\end{figure*}

\begin{figure*}
\centering
\includegraphics[width=\textwidth]{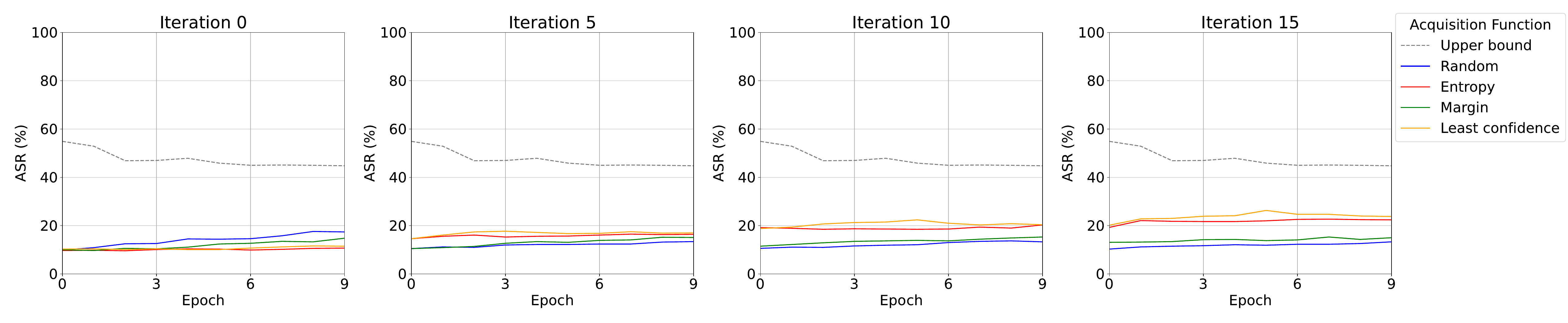} 
\caption{ASR (\%) over training epochs on Fashion-MNIST under the CL trigger with 1.0\% poisoning.}
% \label{fig:asr-cifar-sig}
\end{figure*}

\begin{figure*}
\centering
\includegraphics[width=\textwidth]{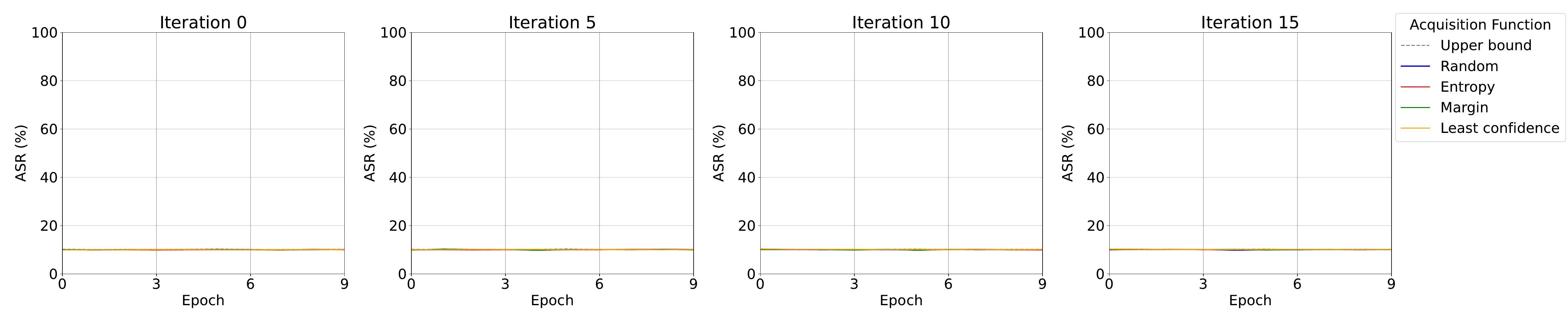} 
\caption{ASR (\%) over training epochs on SVHN under the CL trigger with 1.0\% poisoning.}
% \label{fig:asr-cifar-sig}
\end{figure*}

\begin{figure*}
\centering
\includegraphics[width=\textwidth]{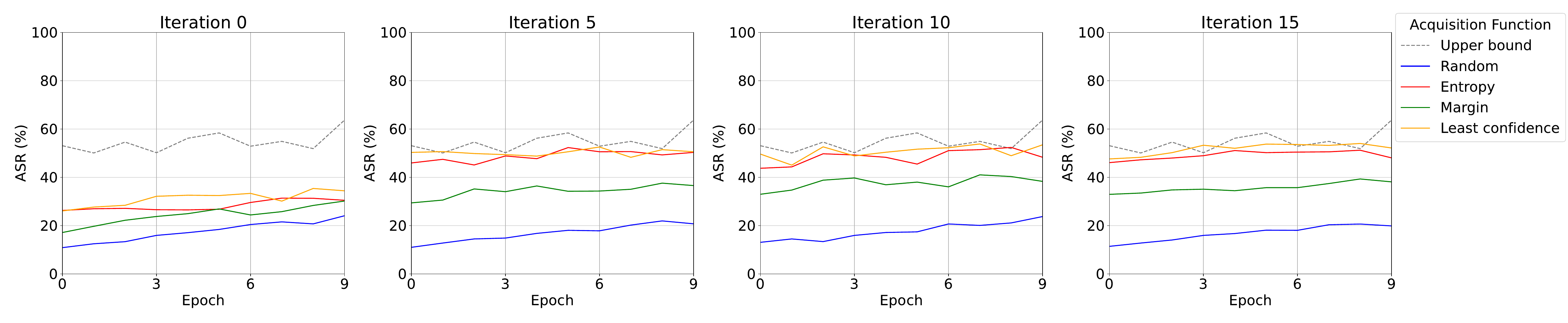} 
\caption{ASR (\%) over training epochs on CIFAR-10 under the SIG trigger with 1.0\% poisoning.}
% \label{fig:asr-cifar-sig}
\end{figure*}

\begin{figure*}
\centering
\includegraphics[width=\textwidth]{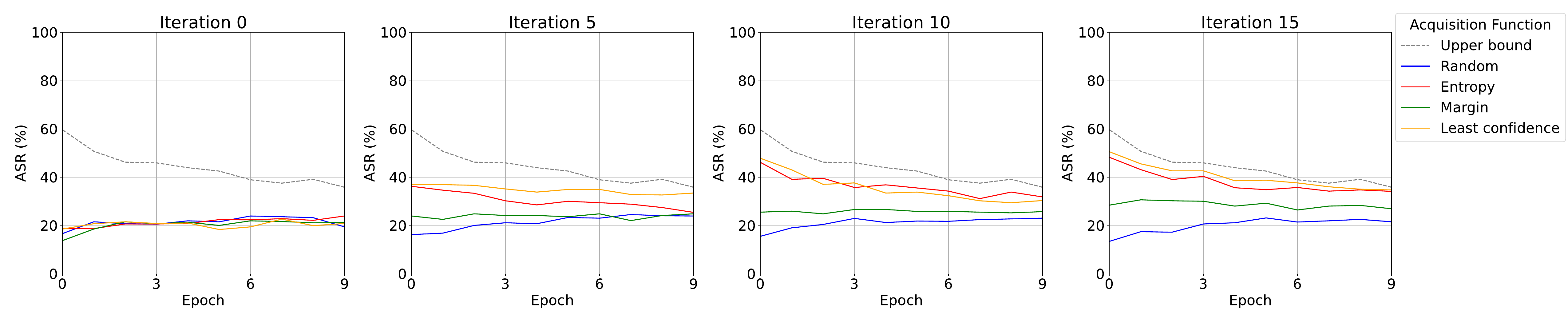} 
\caption{ASR (\%) over training epochs on Fashion-MNIST under the SIG trigger with 1.0\% poisoning.}
% \label{fig:asr-cifar-sig}
\end{figure*}

\begin{figure*}
\centering
\includegraphics[width=\textwidth]{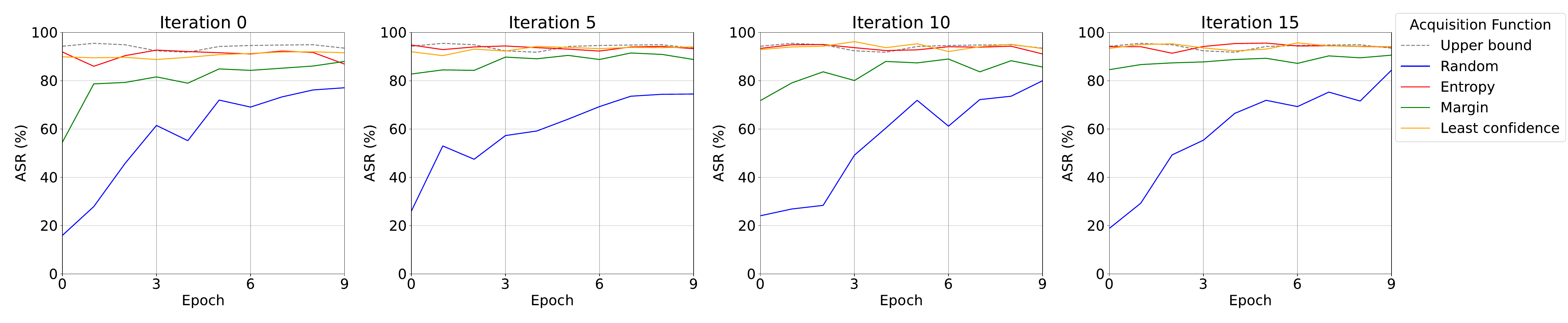} 
\caption{ASR (\%) over training epochs on SVHN under the SIG trigger with 1.0\% poisoning.}
% \label{fig:asr-cifar-sig}
\end{figure*}

\clearpage

\subsection{ASR over different classes}
Figure 17 and Figure 18 show the ASR (\%) of epoch 9 over different classes on Fashion-MNIST and SVHN dataset.
\begin{figure}[ht]
\centering
\includegraphics[width=\columnwidth]{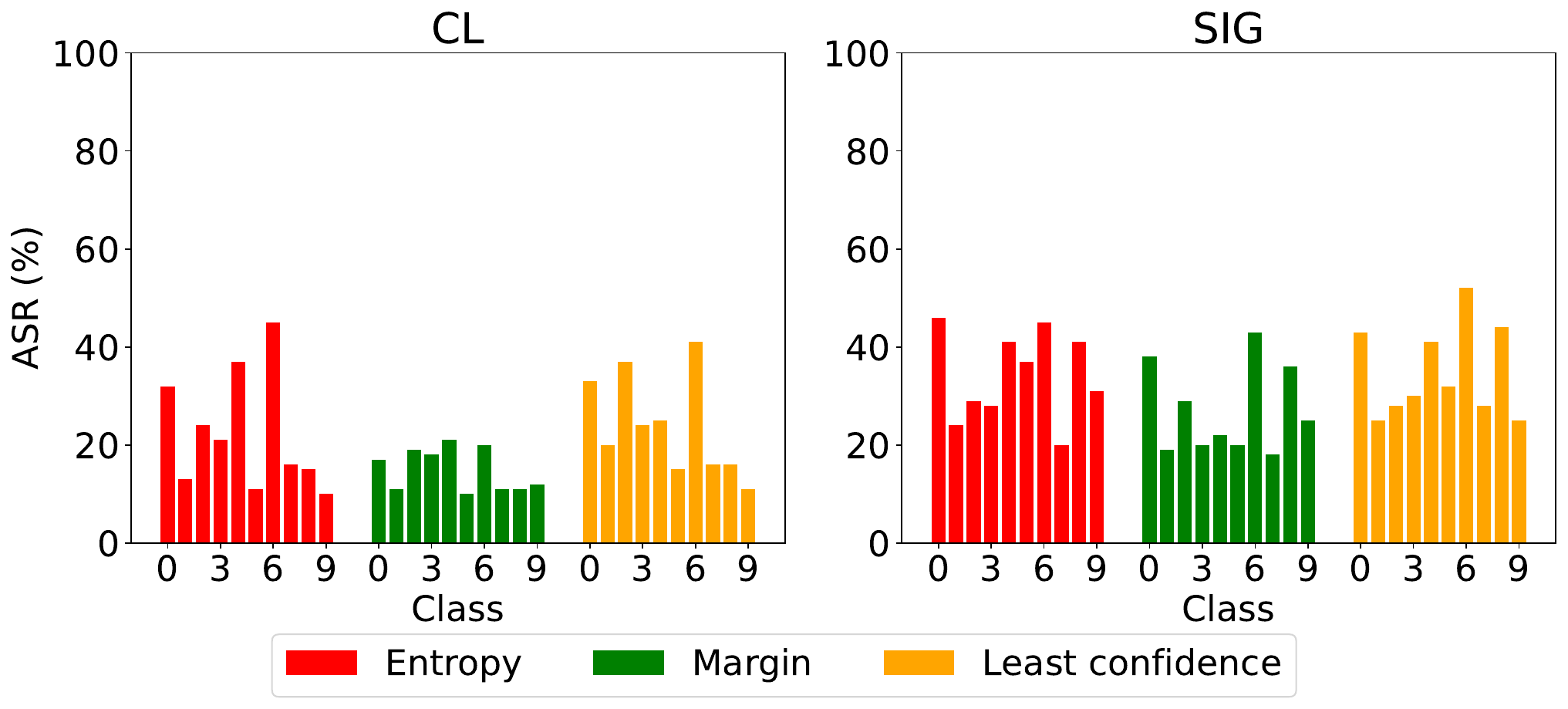} 
\caption{ASR (\%) of epoch 9 over different classes on Fashion-MNIST under the CL and SIG trigger with 1.0\% poisoning and 15 optimization iterations. Colored lines represent different acquisition functions, while the dashed line denotes the estimated upper bound achieved when all poisoned samples are selected in the first AL epoch.}
% \label{fig:class_study}
\end{figure}

\begin{figure}[ht]
\centering
\includegraphics[width=\columnwidth]{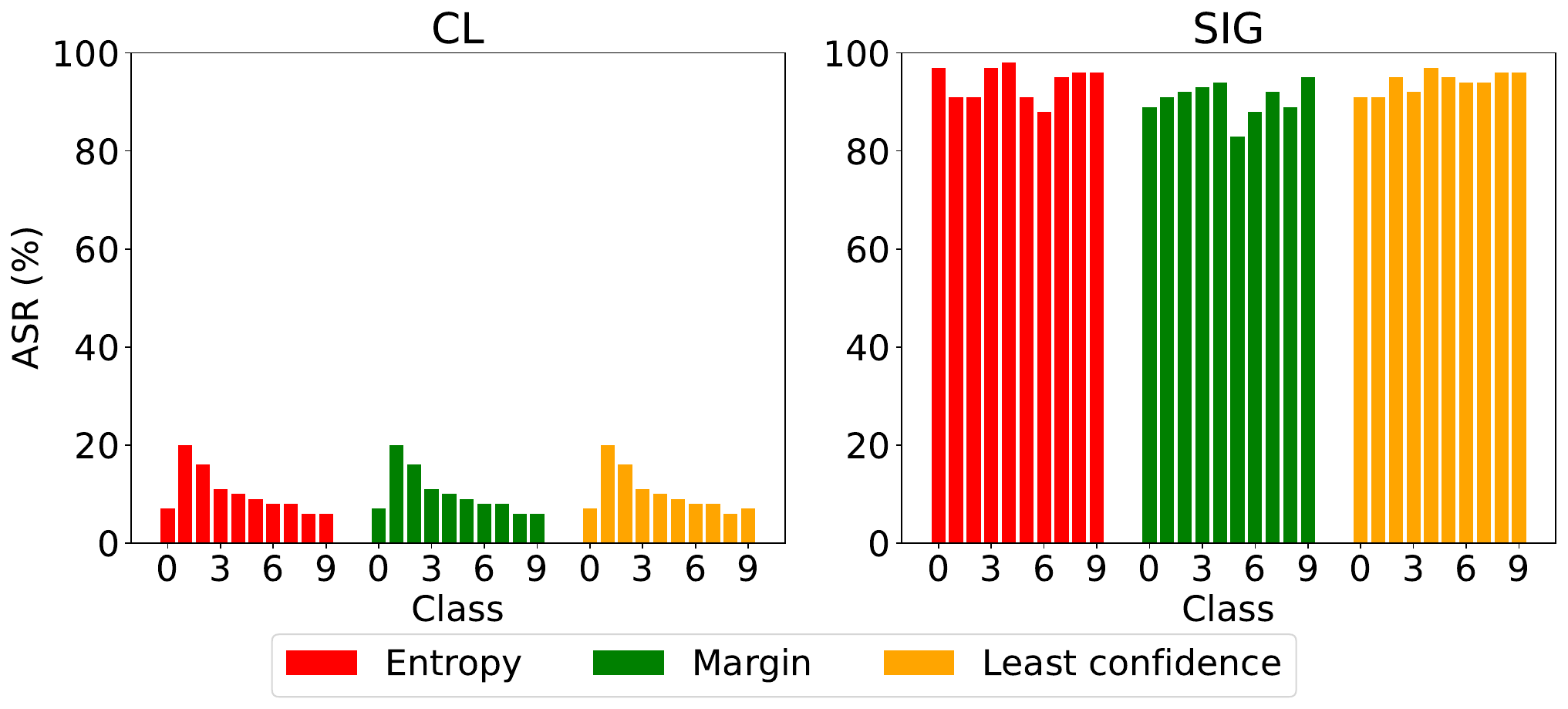} 
\caption{ASR (\%) of epoch 9 over different classes on SVHN under the CL and SIG trigger with 1.0\% poisoning and 15 optimization iterations. Colored lines represent different acquisition functions, while the dashed line denotes the estimated upper bound achieved when all poisoned samples are selected in the first AL epoch.}
% \label{fig:class_study}
\end{figure}

\subsection{OOD data accuracy}
Tables 27–38 show the prediction accuracy of the fine-tuned models on OOD data (i.e., all benign unlabeled data). The results indicate that the backdoor injection process does not affect the model's performance on benign samples.

\begin{table}[ht]
\small
  \centering
  \caption{OOD Accuray (\%) on CIFAR-10 under the CL attack with 0.5\% poisoning for different acquisition functions and optimization rounds. The first row reports the estimated upper bound, where no poisoned samples are injected in unlabeled pool. ACC refers to OOD accuracy and Iter refers to optimization iteration.}
   
   \begin{tabular}{c|cccc}
    \toprule
    ACC (\%) & Iter 0 & Iter 5 & Iter 10 & Iter 15 \\
    \midrule
    Upper bound & 55.0  & 55.0  & 55.0  & 55.0  \\
    Random & 56.3  & 56.5  & 56.3  & 57.5  \\
    Entropy & 55.7  & 56.0  & 55.9  & 55.4  \\
    Margin & 56.9  & 57.0  & 57.9  & 57.5  \\
    Least Confidence & 56.3  & 56.7  & 56.6  & 56.0  \\
    \bottomrule
    \end{tabular}%
    
  % \label{tab:asr-cifar-sig}%
\end{table}%

\begin{table}[ht]
\small
  \centering
  \caption{OOD Accuray (\%) on Fashion-MNIST under the CL attack with 0.5\% poisoning for different acquisition functions and optimization rounds. The first row reports the estimated upper bound, where no poisoned samples are injected in unlabeled pool. ACC refers to OOD accuracy and Iter refers to optimization iteration.}
   
    \begin{tabular}{c|cccc}
    \toprule
    ACC (\%) & Iter 0 & Iter 5 & Iter 10 & Iter 15 \\
    \midrule
    Upper bound & 51.0  & 51.0  & 51.0  & 51.0  \\
    Random & 50.3  & 50.1  & 49.6  & 50.4  \\
    Entropy & 53.3  & 52.7  & 53.2  & 53.0  \\
    Margin & 51.5  & 51.6  & 52.2  & 52.1  \\
    Least Confidence & 52.7  & 52.9  & 52.7  & 53.1  \\
    \bottomrule
    \end{tabular}%
    
  % \label{tab:asr-cifar-sig}%
\end{table}%

\begin{table}[ht]
\small
  \centering
  \caption{OOD Accuray (\%) on SVHN under the CL attack with 0.5\% poisoning for different acquisition functions and optimization rounds. The first row reports the estimated upper bound, where no poisoned samples are injected in unlabeled pool. ACC refers to OOD accuracy and Iter refers to optimization iteration.}
   
   \begin{tabular}{c|cccc}
    \toprule
    ACC (\%) & Iter 0 & Iter 5 & Iter 10 & Iter 15 \\
    \midrule
    Upper bound & 62.0  & 62.0  & 62.0  & 62.0  \\
    Random & 63.0  & 62.7  & 62.6  & 62.3  \\
    Entropy & 63.0  & 62.9  & 61.5  & 61.9  \\
    Margin & 64.4  & 64.0  & 63.6  & 63.9  \\
    Least Confidence & 64.1  & 64.0  & 62.9  & 63.2  \\
    \bottomrule
    \end{tabular}%
    
  % \label{tab:asr-cifar-sig}%
\end{table}%

\begin{table}[ht]
\small
  \centering
  \caption{OOD Accuray (\%) on CIFAR-10 under the SIG attack with 0.5\% poisoning for different acquisition functions and optimization rounds. The first row reports the estimated upper bound, where no poisoned samples are injected in unlabeled pool. ACC refers to OOD accuracy and Iter refers to optimization iteration.}
   
     \begin{tabular}{c|cccc}
    \toprule
    ACC (\%) & Iter 0 & Iter 5 & Iter 10 & Iter 15 \\
    \midrule
    Upper bound & 55.0  & 55.0  & 55.0  & 55.0  \\
    Random & 56.5  & 56.8  & 57.3  & 57.8  \\
    Entropy & 56.3  & 55.8  & 56.1  & 55.2  \\
    Margin & 57.5  & 57.4  & 57.4  & 57.2  \\
    Least Confidence & 56.7  & 57.2  & 56.5  & 56.6  \\
    \bottomrule
    \end{tabular}%
    
  % \label{tab:asr-cifar-sig}%
\end{table}%

\begin{table}[ht]
\small
  \centering
  \caption{OOD Accuray (\%) on Fashion-MNIST under the SIG attack with 0.5\% poisoning for different acquisition functions and optimization rounds. The first row reports the estimated upper bound, where no poisoned samples are injected in unlabeled pool. ACC refers to OOD accuracy and Iter refers to optimization iteration.}
   
    \begin{tabular}{c|cccc}
    \toprule
    ACC (\%) & Iter 0 & Iter 5 & Iter 10 & Iter 15 \\
    \midrule
    Upper bound & 51.0  & 51.0  & 51.0  & 51.0  \\
    Random & 50.2  & 50.5  & 50.6  & 50.1  \\
    Entropy & 53.6  & 52.5  & 52.4  & 52.5  \\
    Margin & 52.3  & 52.1  & 51.9  & 52.2  \\
    Least Confidence & 53.1  & 52.9  & 53.0  & 53.0  \\
    \bottomrule
    \end{tabular}%
    
  % \label{tab:asr-cifar-sig}%
\end{table}%

\begin{table}[ht]
\small
  \centering
  \caption{OOD Accuray (\%) on SVHN under the SIG attack with 0.5\% poisoning for different acquisition functions and optimization rounds. The first row reports the estimated upper bound, where no poisoned samples are injected in unlabeled pool. ACC refers to OOD accuracy and Iter refers to optimization iteration.}

    \begin{tabular}{c|cccc}
    \toprule
    ACC (\%) & Iter 0 & Iter 5 & Iter 10 & Iter 15 \\
    \midrule
    Upper bound & 62.0  & 62.0  & 62.0  & 62.0  \\
    Random & 62.3  & 61.9  & 62.3  & 62.1  \\
    Entropy & 61.2  & 61.5  & 61.3  & 62.1  \\
    Margin & 64.4  & 64.2  & 63.7  & 63.5  \\
    Least Confidence & 62.5  & 61.9  & 62.3  & 62.1  \\
    \bottomrule
    \end{tabular}%
    
  % \label{tab:asr-cifar-sig}%
\end{table}%

\begin{table}[ht]
\small
  \centering
  \caption{OOD Accuray (\%) on CIFAR-10 under the CL attack with 1.0\% poisoning for different acquisition functions and optimization rounds. The first row reports the estimated upper bound, where no poisoned samples are injected in unlabeled pool. ACC refers to OOD accuracy and Iter refers to optimization iteration.}
   
    \begin{tabular}{c|cccc}
    \toprule
    ACC (\%) & Iter 0 & Iter 5 & Iter 10 & Iter 15 \\
    \midrule
    Upper bound & 55.0  & 55.0  & 55.0  & 55.0  \\
    Random & 56.2  & 57.2  & 57.0  & 56.8  \\
    Entropy & 56.2  & 56.2  & 55.4  & 55.4  \\
    Margin & 57.0  & 57.3  & 57.4  & 57.6  \\
    Least Confidence & 56.1  & 56.2  & 55.6  & 55.7  \\
    \bottomrule
    \end{tabular}%
    
  % \label{tab:asr-cifar-sig}%
\end{table}%

\begin{table}[ht]
\small
  \centering
  \caption{OOD Accuray (\%) on Fashion-MNIST under the CL attack with 1.0\% poisoning for different acquisition functions and optimization rounds. The first row reports the estimated upper bound, where no poisoned samples are injected in unlabeled pool. ACC refers to OOD accuracy and Iter refers to optimization iteration.}
   
    \begin{tabular}{c|cccc}
    \toprule
    ACC (\%) & Iter 0 & Iter 5 & Iter 10 & Iter 15 \\
    \midrule
    Upper bound & 51.0  & 51.0  & 51.0  & 51.0  \\
    Random & 49.7  & 50.5  & 49.9  & 50.2  \\
    Entropy & 53.1  & 53.3  & 53.4  & 52.8  \\
    Margin & 51.3  & 52.2  & 52.0  & 52.0  \\
    Least Confidence & 52.8  & 52.7  & 53.0  & 52.4  \\
    \bottomrule
    \end{tabular}%
    
  % \label{tab:asr-cifar-sig}%
\end{table}%

\begin{table}[ht]
\small
  \centering
  \caption{OOD Accuray (\%) on SVHN under the CL attack with 1.0\% poisoning for different acquisition functions and optimization rounds. The first row reports the estimated upper bound, where no poisoned samples are injected in unlabeled pool. ACC refers to OOD accuracy and Iter refers to optimization iteration.}
   
    \begin{tabular}{c|cccc}
    \toprule
    ACC (\%) & Iter 0 & Iter 5 & Iter 10 & Iter 15 \\
    \midrule
    Upper bound & 62.0  & 62.0  & 62.0  & 62.0  \\
    Random & 62.1  & 61.8  & 62.1  & 62.0  \\
    Entropy & 61.3  & 62.0  & 61.9  & 61.1  \\
    Margin & 63.2  & 63.8  & 63.4  & 63.3  \\
    Least Confidence & 62.5  & 62.9  & 63.1  & 62.7  \\
    \bottomrule
    \end{tabular}%
    
  % \label{tab:asr-cifar-sig}%
\end{table}%

\begin{table}[ht]
\small
  \centering
  \caption{OOD Accuray (\%) on CIFAR-10 under the SIG attack with 1.0\% poisoning for different acquisition functions and optimization rounds. The first row reports the estimated upper bound, where no poisoned samples are injected in unlabeled pool. ACC refers to OOD accuracy and Iter refers to optimization iteration.}
   
    \begin{tabular}{c|cccc}
    \toprule
    ACC (\%) & Iter 0 & Iter 5 & Iter 10 & Iter 15 \\
    \midrule
    Upper bound & 55.0  & 55.0  & 55.0  & 55.0  \\
    Random & 57.4  & 58.2  & 57.7  & 57.5  \\
    Entropy & 56.6  & 55.8  & 55.4  & 55.3  \\
    Margin & 57.6  & 57.6  & 57.5  & 57.5  \\
    Least Confidence & 56.9  & 56.0  & 56.0  & 56.1  \\
    \bottomrule
    \end{tabular}%
    
  % \label{tab:asr-cifar-sig}%
\end{table}%

\begin{table}[ht]
\small
  \centering
  \caption{OOD Accuray (\%) on Fashion-MNIST under the SIG attack with 1.0\% poisoning for different acquisition functions and optimization rounds. The first row reports the estimated upper bound, where no poisoned samples are injected in unlabeled pool. ACC refers to OOD accuracy and Iter refers to optimization iteration.}
   
    \begin{tabular}{c|cccc}
    \toprule
    ACC (\%) & Iter 0 & Iter 5 & Iter 10 & Iter 15 \\
    \midrule
    Upper bound & 51.0  & 51.0  & 51.0  & 51.0  \\
    Random & 49.7  & 50.3  & 50.5  & 50.3  \\
    Entropy & 53.5  & 52.7  & 52.6  & 53.1  \\
    Margin & 52.7  & 52.3  & 52.4  & 52.1  \\
    Least Confidence & 53.1  & 53.3  & 52.4  & 52.6  \\
    \bottomrule
    \end{tabular}%
    
  % \label{tab:asr-cifar-sig}%
\end{table}%

\begin{table}[ht]
\small
  \centering
  \caption{OOD Accuray (\%) on SVHN under the SIG attack with 1.0\% poisoning for different acquisition functions and optimization rounds. The first row reports the estimated upper bound, where no poisoned samples are injected in unlabeled pool. ACC refers to OOD accuracy and Iter refers to optimization iteration.}
   
    \begin{tabular}{c|cccc}
    \toprule
    ACC (\%) & Iter 0 & Iter 5 & Iter 10 & Iter 15 \\
    \midrule
    Upper bound & 62.0  & 62.0  & 62.0  & 62.0  \\
    Random & 62.6  & 62.8  & 62.6  & 62.2  \\
    Entropy & 62.3  & 61.2  & 61.7  & 61.1  \\
    Margin & 63.9  & 64.0  & 63.7  & 64.2  \\
    Least Confidence & 62.8  & 62.8  & 62.2  & 62.2  \\
    \bottomrule
    \end{tabular}%
    
  % \label{tab:asr-cifar-sig}%
\end{table}%

% \clearpage
% \subsection{Correlation between $R_{select}$ and ASR}
% Table 37 shows the Poisoned Selection Rate and the corresponding ASR on the CIFAR-10 dataset using the SIG trigger with a poisoning ratio of 0.5\%.
\begin{table*}
\small
  \centering
  \caption{Raw data to calculate correlation between $R_{select}$ and ASR}
    \begin{tabular}{ccccccccccccc}
    \toprule
    \multirow{2}[4]{*}{Iteration} & \multirow{2}[4]{*}{Acquisition Fcuntion} & \multirow{2}[4]{*}{Metric} & \multicolumn{10}{c}{Epoch} \\
\cmidrule{4-13}          &       &       & 0     & 1     & 2     & 3     & 4     & 5     & 6     & 7     & 8     & 9 \\
    \midrule
    \multirow{8}[2]{*}{0} & \multirow{2}[1]{*}{Random} & $R_{select}$ & 11.5  & 12.4  & 12.4  & 12.2  & 12.8  & 14.4  & 13.6  & 15.0  & 15.2  & 14.6 \\
          &       & ASR   & 1.6   & 2.2   & 2.8   & 3.4   & 3.8   & 5.2   & 5.6   & 6.0   & 6.8   & 7.6 \\
          & \multirow{2}[0]{*}{Entropy} & $R_{select}$ & 25.5  & 26.3  & 23.6  & 26.8  & 26.7  & 26.8  & 28.4  & 27.4  & 27.3  & 29.7 \\
          &       & ASR   & 23.8  & 24.6  & 24.9  & 25.6  & 25.9  & 26.3  & 27.2  & 27.9  & 28.5  & 29.1 \\
          & \multirow{2}[0]{*}{Margin} & $R_{select}$ & 13.0  & 14.0  & 16.2  & 17.1  & 17.7  & 20.2  & 20.0  & 21.4  & 21.9  & 21.0 \\
          &       & ASR   & 4.6   & 5.5   & 6.7   & 7.9   & 9.7   & 10.7  & 11.7  & 12.9  & 13.7  & 14.6 \\
          & \multirow{2}[1]{*}{Least Confidence} & $R_{select}$ & 25.3  & 25.5  & 26.5  & 25.4  & 26.8  & 28.3  & 29.4  & 28.1  & 27.7  & 27.1 \\
          &       & ASR   & 23.8  & 24.1  & 24.5  & 24.9  & 25.8  & 26.3  & 27.1  & 27.4  & 28.1  & 28.5 \\
    \midrule
    \multirow{8}[2]{*}{5} & \multirow{2}[1]{*}{Random} & $R_{select}$ & 10.7  & 12.1  & 11.9  & 12.5  & 14.1  & 13.9  & 15.8  & 16.2  & 15.8  & 16.0 \\
          &       & ASR   & 0.6   & 1.6   & 3.4   & 4.4   & 5.0   & 5.8   & 7.4   & 8.0   & 9.2   & 10.0 \\
          & \multirow{2}[0]{*}{Entropy} & $R_{select}$ & 39.1  & 42.4  & 41.8  & 42.3  & 43.1  & 41.4  & 40.2  & 42.5  & 41.8  & 43.6 \\
          &       & ASR   & 75.4  & 75.5  & 75.5  & 75.6  & 75.7  & 75.7  & 75.7  & 75.7  & 75.7  & 75.9 \\
          & \multirow{2}[0]{*}{Margin} & $R_{select}$ & 22.5  & 22.2  & 25.6  & 26.9  & 24.9  & 25.5  & 26.3  & 27.6  & 27.1  & 27.3 \\
          &       & ASR   & 18.0  & 19.2  & 20.0  & 20.5  & 21.1  & 21.5  & 22.3  & 23.1  & 23.7  & 24.3 \\
          & \multirow{2}[1]{*}{Least Confidence} & $R_{select}$ & 43.8  & 41.9  & 42.7  & 44.6  & 43.3  & 44.5  & 42.2  & 44.8  & 45.6  & 43.3 \\
          &       & ASR   & 78.8  & 78.9  & 79.1  & 79.1  & 79.2  & 79.2  & 79.3  & 79.6  & 79.7  & 79.7 \\
    \midrule
    \multirow{8}[2]{*}{10} & \multirow{2}[1]{*}{Random} & $R_{select}$ & 12.3  & 11.0  & 12.4  & 13.5  & 12.7  & 14.2  & 15.3  & 15.4  & 16.8  & 17.7 \\
          &       & ASR   & 1.4   & 2.4   & 3.2   & 3.6   & 4.8   & 5.4   & 6.2   & 7.2   & 8.2   & 9.4 \\
          & \multirow{2}[0]{*}{Entropy} & $R_{select}$ & 41.0  & 43.5  & 42.8  & 40.2  & 43.0  & 42.2  & 42.7  & 46.7  & 42.7  & 43.8 \\
          &       & ASR   & 84.6  & 84.6  & 84.6  & 84.6  & 84.6  & 84.6  & 84.6  & 84.7  & 84.7  & 84.8 \\
          & \multirow{2}[0]{*}{Margin} & $R_{select}$ & 25.6  & 25.5  & 27.2  & 27.7  & 28.3  & 29.4  & 27.4  & 29.8  & 29.7  & 28.5 \\
          &       & ASR   & 26.2  & 26.5  & 26.8  & 27.4  & 27.9  & 28.4  & 28.9  & 29.5  & 29.8  & 30.1 \\
          & \multirow{2}[1]{*}{Least Confidence} & $R_{select}$ & 45.4  & 41.9  & 44.2  & 43.5  & 48.1  & 42.3  & 43.0  & 47.8  & 47.1  & 47.7 \\
          &       & ASR   & 89.4  & 89.4  & 89.4  & 89.4  & 89.4  & 89.4  & 89.4  & 89.5  & 89.5  & 89.5 \\
    \midrule
    \multirow{8}[2]{*}{15} & \multirow{2}[1]{*}{Random} & $R_{select}$ & 11.5  & 11.0  & 12.5  & 12.8  & 15.2  & 15.4  & 14.6  & 14.7  & 16.0  & 16.4 \\
          &       & ASR   & 0.8   & 1.6   & 2.4   & 3.2   & 5.2   & 5.6   & 5.8   & 7.0   & 7.8   & 8.6 \\
          & \multirow{2}[0]{*}{Entropy} & $R_{select}$ & 41.2  & 41.3  & 42.7  & 44.4  & 47.3  & 45.3  & 44.2  & 46.2  & 45.9  & 43.7 \\
          &       & ASR   & 89.4  & 89.4  & 89.4  & 89.4  & 89.4  & 89.4  & 89.4  & 89.4  & 89.4  & 89.4 \\
          & \multirow{2}[0]{*}{Margin} & $R_{select}$ & 29.8  & 32.6  & 28.7  & 30.4  & 32.2  & 34.0  & 33.8  & 33.5  & 33.5  & 30.0 \\
          &       & ASR   & 32.4  & 32.5  & 32.9  & 33.6  & 34.2  & 34.5  & 35.1  & 35.4  & 35.5  & 35.5 \\
          & \multirow{2}[1]{*}{Least Confidence} & $R_{select}$ & 43.7  & 45.8  & 45.8  & 47.6  & 47.0  & 48.8  & 47.9  & 47.2  & 46.2  & 48.2 \\
          &       & ASR   & 95.8  & 95.8  & 95.8  & 95.8  & 95.8  & 95.8  & 95.8  & 95.8  & 95.8  & 95.8 \\
    \bottomrule
    \end{tabular}%
  % \label{tab:addlabel}%
\end{table*}%

\newpage
\end{document}